\documentclass[showpacs,preprintnumbers,amsmath,amssymb,nofootinbib,superscriptaddress,floatfix,twocolumn]{revtex4}
\usepackage[normalem]{ulem}
\usepackage{graphicx}
\usepackage{dcolumn}
\usepackage{bm}
\usepackage{color}
\usepackage[colorlinks,citecolor=blue,urlcolor=blue,linkcolor=blue]{hyperref}
\usepackage{multirow}
\usepackage{subfigure}

\newcommand{\f}{\frac}
\newcommand{\lt}{\left}
\newcommand{\n}{\nonumber}
\newcommand{\p}{\partial}
\newcommand{\dd}{{\rm d}}
\newcommand{\rt}{\right}

\newcommand{\arxgr}[1]{\href{http://arxiv.org/abs/#1}{{\ttfamily arXiv:#1[gr-qc]}}}
\newcommand{\arxth}[1]{\href{http://arxiv.org/abs/#1}{{\ttfamily arXiv:#1[hep-th]}}}
\newcommand{\arxas}[1]{\href{http://arxiv.org/abs/#1}{{\ttfamily arXiv:#1[astro-ph]}}}
\newcommand{\Arxgr}[1]{\href{http://arxiv.org/abs/gr-qc/#1}{{\ttfamily arXiv:#1[gr-qc]}}}
\newcommand{\Arxth}[1]{\href{http://arxiv.org/abs/hep-th/#1}{{\ttfamily arXiv:#1[hep-th]}}}

\begin{document}
\title{Hawking--Page phase transitions in four-dimensional Einstein--Gauss--Bonnet gravity}

\author{Yuan-Yuan Wang}
%\email{a@mail.neu.edu.cn}
\affiliation{Department of Physics, College of Sciences, Northeastern University, Shenyang 110819, China}
\affiliation{Department of Physics, School of Physics, Peking University, Beijing 100871, China}
\author{Bing-Yu Su}
%\email{20151150@stu.neu.edu.cn}
\affiliation{Department of Physics, College of Sciences, Northeastern University, Shenyang 110819, China}
\author{Nan Li}
\email{linan@mail.neu.edu.cn}
\affiliation{Department of Physics, College of Sciences, Northeastern University, Shenyang 110819, China}
% \date{Received: date / Accepted: date}

\begin{abstract}
The Hawking--Page (HP) phase transitions of the anti-de Sitter black holes in the extended phase space are studied in a novel four-dimensional Einstein--Gauss--Bonnet (4EGB) gravity, which is proposed by rescaling the Gauss--Bonnet (GB) coupling constant $\alpha\to\alpha/(d-4)$ in $d$ dimensions and redefining the four-dimensional gravity in the limit $d \to 4$. The GB term shows nontrivial contributions to both black hole mass and entropy simultaneously, and decreases the HP phase transition temperature $T_{\rm HP}$. Moreover, the HP phase transitions can occur only within a range of pressure in the 4EGB gravity. For the charged black holes, $T_{\rm HP}$ also decreases with the electric potential in the grand canonical ensemble. A general discussion of the HP phase transitions in the Einstein, GB, and 4EGB gravities is also presented.
\end{abstract}
\maketitle

\section{Introduction}

Black hole physics is one of the central topics in modern physics, as it exhibits that a black hole should be considered as a complicated system with temperature and entropy \cite{Bekenstein}. The establishment of the four laws of black hole thermodynamics clearly indicates the profound relationship of thermodynamics, classical gravity, and quantum mechanics, and will help to our final understanding of quantum gravity \cite{law}.

In the last decade, black hole thermodynamics in the extended phase space has attracted increasing attentions \cite{KM}. The basic motivation is to construct the effective black hole volume $V$ and pressure $p$ and to restore the $p$--$V$ term that is absent in black hole thermodynamics. This purpose can be realized in the anti-de Sitter (AdS) space with a negative cosmological constant $\Lambda$. Since the gravitational potential in the AdS space increases at large distances, it offers a positive pressure $p$. If $\Lambda$ is further allowed to run, $p$ will become a varying thermodynamic quantity, and the effective black hole volume $V$ can be defined as the conjugate variable of $p$. By this means, the missing $p$--$V$ term reappears in the first law of black hole thermodynamics, but in the form $V\,\dd p$, not the usual work term $-p\,\dd V$. Therefore, the black hole mass should not be identified as internal energy, but rather as enthalpy. In the framework of the extended phase space, an AdS black hole shows significant similarities to a non-ideal fluid, especially in their phase transitions and critical phenomena. These remarkable observations have induced a large number of relevant works. See Ref. \cite{rev3} for the recent progresses and references.

The famous Hawking--Page (HP) phase transition possesses an important position among the black hole phase transitions \cite{HP}. It was first studied between the Schwarzschild--AdS black hole and the thermal AdS background, and then extended to the charged AdS [i.e., Reissner--Nordstr\"{o}m--AdS (RN--AdS)] black hole \cite{Chamblin1}. Moreover, it was also interpreted as a confinement--deconfinement phase transition of gauge fields in the AdS/CFT duality \cite{Witten}. In general, the partition function of the black hole--thermal AdS system is dominated by the thermal AdS phase or the black hole phase at low or high temperatures respectively. Therefore, above the phase transition temperature $T_{\rm HP}$, the thermal AdS gas will collapse into a black hole via a first-order phase transition. The HP phase transition has been widely investigated in the literature, with emphasis in the extended phase space \cite{Spallucci:2013jja, Altamirano:2014tva, Xu:2015rfa, Maity:2015ida, Liu:2016uyd, Sahay:2017hlq, Mbarek:2018bau, sanren, HPwo, Wang:2020hjw}. Currently, the research topics are mainly focused in various modified gravity theories.

One of the important modified gravity theories is the Gauss--Bonnet (GB) gravity, which is a special case of the most general Lovelock gravity, with the GB term ${\cal G}$ being the leading-order correction to the Einstein--Hilbert action and including the Einstein gravity as the low energy and small curvature limit. Albeit ${\cal G}$ is quadratic in curvature tensors, the equations of gravitational field are still of second-order and thus avoid ghosts automatically. However, the studies of the GB gravity are usually restricted to high-dimensional physics, as in a four-dimensional manifold, the GB term reduces to a topological invariant and its integral just corresponds to the Euler characteristic of the manifold. From this point of view, the GB term cannot influence space-time structure, global charges, and their conjugate potentials, so it has no dynamical effect and is thus always disregarded in four dimensions, unless coupled to other matter fields. The explorations of the GB gravity in the extended phase space can be found in Refs. \cite{Cai:2013qga, Xu:2013zea, Wei:2014hba, Mo:2014mba, Zou:2014mha, Belhaj:2014eha, Hendi:2016njy, Ghaffarnejad:2018tpr, Mahish:2019tgv, Zeng:2019hux, Hyun:2019gfz, Wei:2019ctz, Ghosh:2019pwy, Haroon:2020vpr, Ye:2020aln}.

Recently, a novel four-dimensional Einstein--Gauss--Bonnet (4EGB) gravity theory was proposed by Glavan and Lin and was applied to maximally symmetric space-time, spherically symmetric black hole, and cosmology \cite{4EGB}. Their basic idea was to rescale the GB coupling constant $\alpha\to\alpha/(d-4)$ in $d$-dimensional space-time and to redefine four-dimensional gravity in the limit $d\to4$. With this rescaling, the GB term was shown to bypass the Lovelock theorem, to be free from the Ostrogradsky instability, and to contribute nontrivial effects to the gravitational dynamics even in four dimensions. This work rapidly aroused intensive research interests, such as black hole solutions \cite{Fernandes, Konoplya:2020qqh, Casalino:2020kbt, Kumar:2020owy, Ghosh:2020syx, Fernandes:2020nbq, Yang:2020jno, Liu:2020yhu}, relativistic stars \cite{Doneva:2020ped, Banerjee:2020yhu}, gravitational lensing \cite{Islam:2020xmy, Jin:2020emq}, stability \cite{Konoplya:2020juj, Zhang:2020sjh, Liu:2020evp, Ge:2020tid}, radiation and accretion \cite{Zhang:2020qam, Liu:2020vkh}, cosmic censorship conjecture \cite{Yang:2020czk, Ying:2020bch}, quasi-normal modes \cite{Konoplya:2020bxa, Aragon:2020qdc}, black hole shadows \cite{Guo:2020zmf, Wei:2020ght, Zeng:2020dco}, thermodynamics and phase transitions \cite{Hegde:2020xlv, Singh:2020xju, HosseiniMansoori:2020yfj, Wei:2020poh, EslamPanah:2020hoj, Qiao:2020hkx, Li:2020vpo}, gravitational waves \cite{Odintsov:2020zkl, Odintsov:2020sqy, Odintsov:2020xji, Oikonomou:2020oil}, cosmological applications \cite{Li:2020tlo, Haghani:2020ynl, Narain:2020qhh, Aoki:2020iwm, MohseniSadjadi:2020jmc}, and observational constraints \cite{Clifton:2020xhc, Feng:2020duo, Garcia-Aspeitia:2020uwq}. Meanwhile, it also received some criticisms \cite{Ai:2020peo, Gurses:2020ofy, Mahapatra:2020rds, Shu:2020cjw, Bonifacio:2020vbk, Hennigar:2020lsl, Tian:2020nzb, Arrechea:2020evj} (e.g., the vacuum is not well-defined), and several variants of Ref. \cite{4EGB} have already been constructed to deal with these troubles \cite{Aoki:2020lig, Easson:2020mpq, Lin:2020kqe}. Nevertheless, at present it is well worth investigating the various applications of this 4EGB gravity and exploring its relevant effects to the most extent.

The basic purpose of this paper is to study the HP phase transitions in the extended phase space in the 4EGB gravity. This paper is a successive research of our previous work \cite{HPwo}, in which we investigated the HP phase transitions of four-dimensional AdS black holes in the Einstein and GB gravities,\footnote{Attention, in four-dimensional GB gravity, although the GB term has no effect on gravitational dynamics, it does influence the black hole entropy and is thus still physically meaningful.} and found that the HP temperature decreases at large electric potentials and angular velocities and also decreases with the GB coupling constant. The main improvements in our present work are threefold. First, in the GB gravity, only the black hole entropy receives a shift from the GB term, but the black hole mass is unaltered. However, in the 4EGB gravity, both the black hole entropy and mass are affected by the GB term, as it influences black hole thermodynamics and dynamics simultaneously. Second, the correction to the black hole entropy is more complicated in the 4EGB gravity than the GB gravity. Third, in the Einstein and GB gravities, the minimal black hole temperature is always positive, but in the 4EGB gravity, it is allowed to reach zero. All these ingredients naturally lead to the evident similarities and also distinct dissimilarities among the Einstein, GB, and 4EGB gravities in the HP phase transitions, and we will systematically consider all these issues in this paper.

This paper is organized as follows. In Sect. \ref{sec:BHT}, we explain the 4EGB gravity and discuss the extended phase space and the HP phase transition in more detail. In Sects. \ref{sec:HPSAdS} and \ref{sec:HPRNAdS}, the HP phase transitions of the Schwarzschild--AdS and RN--AdS black holes are studied in the 4EGB gravity in order. We conclude in Sect. \ref{sec:con}. In this paper, we work in the natural system of units and set $c=G_{\rm N}=\hbar=k_{\rm B}=1$, but the $d$-dimensional gravitational constant $G_d$ is kept without loss of generality. %For the RN--AdS black holes, the grand canonical ensemble with fixed electric potential is adopted.

\section{Black hole thermodynamics in the 4EGB gravity} \label{sec:BHT}

In this section, in the framework of the 4EGB gravity, we outline the thermodynamic properties of the RN--AdS black holes in the extended phase space and discuss the HP phase transition in more detail.

\subsection{Novel 4EGB gravity} \label{sec:4EGB}

We start from the action of the gravitational field of $d$-dimensional GB gravity \cite{lue},
\begin{flalign}
&S_d=\f{1}{16\pi G_d}\int\dd^d x\,\sqrt{-g}(R-2\Lambda+\alpha{\cal G}), &\label{action}
\end{flalign}
where $\alpha$ is the GB coupling constant, and ${\cal G}$ is the GB term,
\begin{flalign}
&{\cal G}:=R_{\mu\nu\lambda\rho}R^{\mu\nu\lambda\rho}-4R_{\mu\nu}R^{\mu\nu}+R^2, &\n
\end{flalign}
where $R_{\mu\nu\lambda\rho}$ is the Riemann tensor, $R_{\mu\nu}$ is the Ricci tensor, and $R$ is the Ricci scalar. When we go to four dimensions, we may parameterize the $d$-dimensional metric via the Kaluza--Klein reduction ansatz as $\dd s^2_d=\dd s^2_4+e^{2\phi}\,\dd s^2_{d-4}$, where $\phi$ is a scalar field (a metric function depending only on the external four-dimensional coordinates), $\dd s^2_4$ is the four-dimensional line element, and $\dd s^2_{d-4}$ is the line element of the internal maximally symmetric space of $(d-4)$ dimensions (we focus on the flat internal space for simplicity). By this means, the $d$-dimensional GB gravity can be compactified on the $(d-4)$-dimensional internal space, and the action $S_d$ reduces to
\begin{flalign}
&\f{1}{16\pi}\int\dd^4x\,\sqrt{-g}\Big\{R-2\Lambda+(d-4)(d-5)(\p\phi)^2 &\n\\
& +\alpha\Big[{\cal G}-4(d-4)(d-5)G^{\mu\nu}\p_\mu\phi\p_\nu\phi &\n\\
& -(d-4)(d-5)(d-6)\Big(2(\p\phi)^2\p^2\phi+(d-5)((\p\phi)^2)^2\Big) \Big]\Big\}, &\n
\end{flalign}
where $G^{\mu\nu}$ is the Einstein tensor. Furthermore, in four dimensions, since the GB term is purely topological, we need to add another term into the action,
\begin{flalign}
&-\f{1}{16\pi}\int\dd^4 x\,\sqrt{-g} \alpha{\cal G}. & \n
\end{flalign}
Next, if the GB coupling constant $\alpha$ is rescaled as
\begin{flalign}
&\alpha\to\f{\alpha}{d-4}, & \n
\end{flalign}
we finally obtain the action of the 4EGB gravity in the limit $d\to 4$ \cite{lue},
\begin{flalign}
&S_4=\f{1}{16\pi}\int\dd^4x\,\sqrt{-g}\lt\{R-2\Lambda+\alpha[\phi{\cal G} +4G^{\mu\nu}\p_\mu\phi\p_\nu\phi\rt. &\n\\
&\quad\quad \lt.-4(\p\phi)^2\p^2\phi+2((\p\phi)^2)^2]\rt\}. \n & %\label{1}
\end{flalign}
Here, we clearly observe that there are non-minimal coupling of the scalar field $\phi$ to the GB term ${\cal G}$ and also the nonlinear self-interacting terms for $\phi$. The same result of $S_4$ has also been obtained in Ref. \cite{Hennigar:2020lsl} by two different methods: the dimensional reduction method and the conformal transformation approach. In this way, the GB term will have nontrivial effect on gravitational dynamics even in four dimensions.

For the black hole solution in the 4EGB gravity, the static and spherically symmetric RN--AdS black hole metric reads
\begin{flalign}
&\dd s^2=-f(r)\,\dd t^2+\f{\dd r^2}{f(r)}+r^2(\dd\theta^2+\sin^2\theta\,\dd\phi^2), &\n
\end{flalign}
and the solution is \cite{Fernandes}
\begin{flalign}
&f(r)=1+\f{r^2}{2\alpha}\lt[1\pm\sqrt{1+4\alpha\lt(\f{2M}{r^3}-\f{Q^2}{r^4}+\f{\Lambda}{3}\rt)}\rt], \label{fr}&
\end{flalign}
where $M$ and $Q$ are the black hole mass and electric charge. Below, we only choose the minus sign in front of the square root. Thus, when $\alpha\to 0$, $f(r)$ reduces to the RN--AdS black hole solution in the Einstein gravity,
\begin{flalign}
&f(r)\to 1-\f{2M}{r}+\f{Q^2}{r^2}-\f{\Lambda r^2}{3}. &\label{RN}
\end{flalign}

\subsection{Black hole thermodynamics in the extended phase space} \label{sec:thermo}

From Eq. (\ref{fr}), the event horizon radius $r_+$ of the RN--AdS black hole can be determined as the largest root of $f(r)=0$. Then, the black hole mass can be extracted as
\begin{flalign}
&M=\f{3r_+^2+3Q^2+3\alpha+8\pi pr_+^4}{6r_+}=M(r_+,p,Q,\alpha), &\label{M}
\end{flalign}
where $p$ is the effective thermodynamic pressure in the extended phase space,
\begin{flalign}
&p=-\f{\Lambda}{8\pi}=\f{3}{8\pi l^2}, &\n
\end{flalign}
with $l$ being the AdS curvature radius. In the limit of vanishing charge and pressure, from Eq. (\ref{M}), we have $r_+=M+\sqrt{M^2-\alpha}$, and this sets an upper bound of $\alpha$ as $\alpha<M^2$. Moreover, as in extra-dimensional physics, $\alpha$ is proportional to the inverse string tension with positive coefficient \cite{Boulware:1985wk}, we set the lower bound of $\alpha$ to be 0 in this work for simplicity. For the discussion with a negative $\alpha$, see Ref. \cite{Guo:2020zmf}.

Next, the Hawking temperature of the RN--AdS black hole is
\begin{flalign}
&T=\f{f'(r_+)}{4\pi}=\f{r_+^2-Q^2-\alpha+8\pi p r_+^4}{4\pi r_+(r_+^2+2\alpha)}=T(r_+,p,Q,\alpha). &\label{T}
\end{flalign}
By rewriting Eq. (\ref{T}), we can establish the equation of state of the RN--AdS black hole in the extended phase space, like that of a non-ideal fluid,
\begin{flalign}
&p=\f{4\pi r_+(r_+^2+2\alpha)T-r_+^2+Q^2+\alpha}{8\pi r_+^4}. &\label{eos}
\end{flalign}

Furthermore, the entropy of the RN--AdS black hole can be calculated as \cite{caii}
\begin{flalign}
&S=\int^{r_+}\f 1T\lt(\f{\p M}{\p r_+}\rt)_{p,Q,\alpha}\,\dd r_+ &\n\\
&\ \ =\pi r_+^2+4\pi\alpha\ln\f{r_+}{L_0}=S(r_+,\alpha), &\label{SS}
\end{flalign}
where $L_0$ is an integral constant to be determined later. Hence, the black hole entropy receives a logarithmic correction in the 4EGB gravity. If we solve $r_+=r_+(S,\alpha)$ from Eq. (\ref{SS}) and substitute it into Eq. (\ref{M}), we can reexpress the RN--AdS black hole mass in terms of the thermodynamic variables as $M=M(r_+(S,\alpha),p,Q,\alpha)$.

A straightforward differentiation of $M$ yields the first law of black hole thermodynamics in the extended phase space,
\begin{flalign}
&\dd M=T\,\dd S+V\,\dd p+\Phi\,\dd Q+X\,\dd \alpha, &\label{first}
\end{flalign}
where $T$, $V$, $\Phi$, and $X$ are the Hawking temperature, thermodynamic volume, electric potential at event horizon, and conjugate variable of $\alpha$ respectively,
\begin{flalign}
&T=\lt(\f{\p M}{\p S}\rt)_{p,Q,\alpha}=\f{(\p M/\p r_+)_{p,Q,\alpha}}{(\p S/\p r_+)_{\alpha}} &\n\\
&\ \ =\f{r_+^2-Q^2-\alpha+8\pi p r_+^4}{4\pi r_+(r_+^2+2\alpha)}=\f{f'(r_+)}{4\pi}, &\label{TT}\\
&V=\lt(\f{\p M}{\p p}\rt)_{S,Q,\alpha}=\lt(\f{\p M}{\p p}\rt)_{r_+,Q,\alpha}=\f{4\pi r_+^3}{3}, &\label{V}\\
&\Phi=\lt(\f{\p M}{\p Q}\rt)_{S,p,\alpha}=\lt(\f{\p M}{\p Q}\rt)_{r_+,p,\alpha}=\f{Q}{r_+},&\label{Phi}\\
&X=\lt(\f{\p M}{\p\alpha}\rt)_{S,p,Q}=\f{(\p M/\p r_+)_{p,Q,\alpha}}{(\p \alpha/\p r_+)_{S}}+\lt(\f{\p M}{\p\alpha}\rt)_{r_+,p,Q} &\n\\
&\ \ =\f{r_+^2-Q^2-\alpha+8\pi p r_+^4}{r_+(r_+^2+2\alpha)}\lt[\lt(\f{\p\ln L_0}{\p\ln\alpha}\rt)_S-\ln\f{r_+}{L_0}\rt]+\f{1}{2r_+} &\n\\
&\ \ =4\pi T\lt[\lt(\f{\p\ln L_0}{\p\ln\alpha}\rt)_S-\ln\f{r_+}{L_0}\rt]+\f{1}{2r_+}. &\label{X}
\end{flalign}

Substituting Eqs. (\ref{TT})--(\ref{X}) into the Smarr relation (i.e., the Gibbs--Duhem relation in traditional thermodynamics),
\begin{flalign}
&M=2TS-2pV+\Phi Q+2X\alpha, &\n
\end{flalign}
in order to guarantee the consistency of this relation and the first law of black hole thermodynamics, we need
\begin{flalign}
&\lt(\f{\p\ln L_0}{\p\ln\alpha}\rt)_S=\f 12. &\n
\end{flalign}
This condition fixes the integral constant $L_0$ in Eq. (\ref{SS}) as $L_0=\sqrt\alpha$, so we finally arrive at the RN--AdS black hole entropy in the 4EGB gravity \cite{Wei:2020poh, lue},
\begin{flalign}
&S =\pi r_+^2+4\pi\alpha\ln\f{r_+}{\sqrt\alpha}. &\label{S}
\end{flalign}
This result can also be obtained via the Iyer--Wald formula \cite{Wald}.\footnote{We show this explicitly in the appendix.}

From Eqs. (\ref{M}) and (\ref{S}), we find that, in the 4EGB gravity, the GB term influences both the black hole mass and entropy. This is more complicated than the case in the GB gravity, in which the GB term merely modifies the black hole entropy by a shift $4\pi\alpha$. Consequently, the HP phase transitions in the 4EGB gravity will exhibit richer thermodynamic behaviors than those in the GB gravity.

\subsection{HP phase transition} \label{sec:HP}

Because the $p$--$V$ term in Eq. (\ref{first}) is $V\,\dd p$, the distinct character of black hole thermodynamics in the extended phase space is that the black hole mass should be identified as enthalpy. Therefore, the corresponding thermodynamic potential of interest should be the Gibbs free energy. Furthermore, since the thermal AdS background is neutral, it is impossible for a black hole with fixed charge $Q$ to undergo the HP phase transition due to the conservation of charge, so all the relevant discussions must be performed in a grand canonical ensemble with fixed electric potential $\Phi$ ($Q$ is allowed to vary). In the grand canonical ensemble, the Gibbs free energy $G$ of the RN--AdS black hole should be constructed as
\begin{flalign}
&G =M-TS-\Phi Q. &\n
\end{flalign}

Substituting Eqs. (\ref{M}), (\ref{T}), (\ref{Phi}), and (\ref{S}) into the above equation, we have
\begin{flalign}
&G=\f{3(1-\Phi^2)r_+^2+3\alpha+8\pi pr_+^4}{6r_+} \n\\
&\quad\quad -\f{(1-\Phi^2)r_+^2-\alpha+8\pi p r_+^4}{4 r_+(r_+^2+2\alpha)}\lt(r_+^2+4\alpha\ln\f{r_+}{\sqrt\alpha}\rt) \n\\
&\ \ =G(r_+,p,\Phi,\alpha). &\label{G}
\end{flalign}
Then, we may solve $r_+=r_+(T,p,\Phi,\alpha)$ from Eq. (\ref{T}), substitute the result into Eq. (\ref{G}), and eventually reexpress the Gibbs free energy in its usual form,
\begin{flalign}
&G=G(T,p,\Phi,\alpha). &\n
\end{flalign}
The analytical solution of $G(T,p,\Phi,\alpha)$ does exist in principle, but we omit it here for its unnecessary lengthy expression.

Due to the Hawking radiation, a stable large black hole (with positive heat capacity and large event horizon radius) can exchange energy and establish the equilibrium with the thermal AdS background. In the black hole--thermal AdS system, first, the Gibbs free energy of the thermal AdS background is zero, since the total number of thermal gas particles is not conserved; second, the Gibbs free energy of the RN--AdS black hole will be shown to decrease with temperature. Hence, below or above the HP temperature, the thermal AdS phase or the black hole phase with the global minimum of Gibbs free energy is thermodynamically preferred respectively, so the criterion of the HP phase transition is
\begin{flalign}
&G=0, &\label{panju}
\end{flalign}
and the HP temperature $T_{\rm HP}$ can be fixed accordingly.

\section{HP phase transitions of the Schwarzschild--AdS black holes} \label{sec:HPSAdS}

In this section, we discuss the HP phase transitions of the Schwarzschild--AdS black holes in the extended phase space. Our discussions consist of two steps: first, to determine the HP temperature $T_{\rm HP}$ as a function of pressure $p$; second, to determine the global phase structure of the black hole--thermal AdS system (i.e., to figure out the dependence of Gibbs free energy $G$ on temperature $T$). For comparison, we study the relevant issues both in the Einstein and 4EGB gravities. In the Einstein gravity, the detailed results can be found in our previous work \cite{HPwo}, and we only refer to them when needed. Below, we focus on the 4EGB gravity and explore its corrections to the Einstein gravity.

\subsection{Einstein gravity} \label{sec:SAdSa0}

We start our discussions in the Einstein gravity with $\alpha=0$. The intermediate variable in calculations in Ref. \cite{HPwo} is the black hole entropy $S$, but now is the event horizon radius $r_+$. These two calculational methods are equivalent when $\alpha=0$. However, when $\alpha\neq0$, the procedure in Ref. \cite{HPwo} will intrinsically not work, as we see from Eq. (\ref{S}) that the dependence of $S$ on $r_+$ is not polynomial, while all the expressions in Eqs. (\ref{M}), (\ref{T}), and (\ref{G}) are written in terms of $r_+$. Therefore, we still need to briefly repeat the calculations here.

In the Einstein gravity, from Eqs. (\ref{T}) and (\ref{G}), the Schwarzschild--AdS black hole temperature and Gibbs free energy reduce to
\begin{flalign}
%M&=\f{3r_++8\pi pr_+^3}{6}, \label{MSAdSwithout}\\
&T=\f{1+8\pi pr_+^2}{4\pi r_+}, &\label{TSAdSwithout}\\
&G=\f{3r_+ -8\pi pr_+^3}{12}.&\label{GSAdSwithout}
\end{flalign}
Hence, at the HP phase transition point, from the criterion in Eq. (\ref{panju}), we obtain $r_+=\sqrt{3/(8\pi p)}$. Substituting it into Eq. (\ref{TSAdSwithout}), the HP temperature is
\begin{flalign}
&T_{\rm HP}=\sqrt{\f{8p}{3\pi}}. &\label{TpSAdSwithout}
\end{flalign}
Actually, the $T_{\rm HP}$--$p$ curve is just the coexistence line in the phase diagram. More importantly, as $p$ has no terminal point in Eq. (\ref{TpSAdSwithout}), the HP phase transition can occur at all pressures, without a critical point, so it is more like a solid--liquid phase transition, rather than a liquid--gas one \cite{Kubiznak:2014zwa}.

\begin{figure}[h]
\begin{center}
\includegraphics[width=0.45\textwidth]{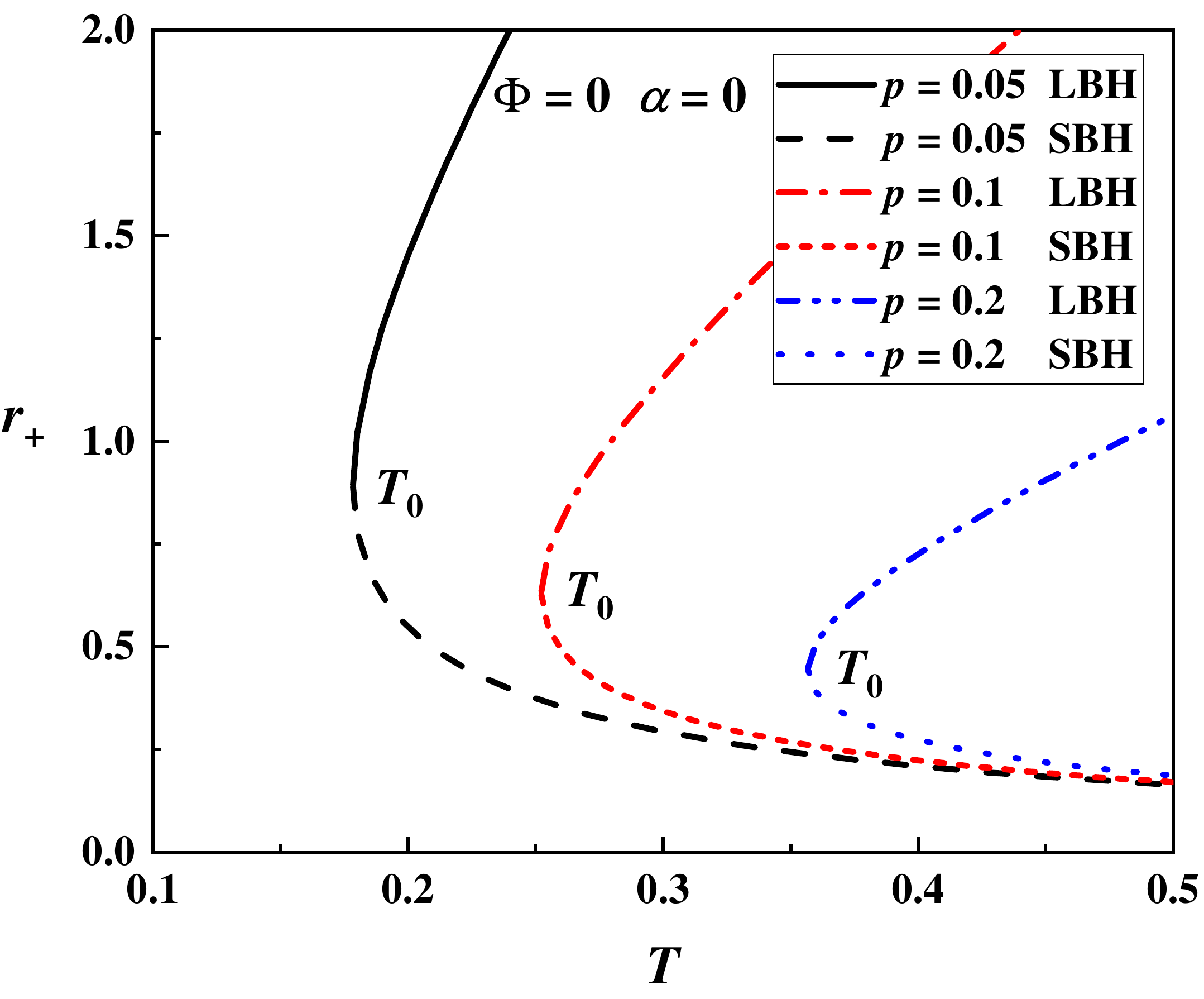}
\end{center}
\caption{The event horizon radii of the Schwarzschild--AdS black holes as a function of temperature (LBH and SBH stand for large and small black holes respectively). At a given pressure $p$, the black hole temperature has a positive minimum $T_0=\sqrt{2p/\pi}$. The large black holes are thermodynamically stable, as they have positive $r_+$--$T$ slopes and positive heat capacities.} \label{f:rTSAdSwithout}
\end{figure}

Then, from Eq. (\ref{TSAdSwithout}), we can solve $r_+$ in terms of $T$ and $p$,
\begin{flalign}
&r_+(T,p)=\f{1}{4\pi p}\lt(\pi T \pm \sqrt{\pi^2 T^2-2\pi p}\rt), &\label{rTSAdSwithout}
\end{flalign}
where $\pm$ correspond to large and small black holes with different event horizon radii respectively. Moreover, from Eq. (\ref{rTSAdSwithout}), the Schwarzschild--AdS black hole must have a positive minimal temperature as
\begin{flalign}
&T_0=\sqrt{\f{2p}{\pi}}. &\n
\end{flalign}
In fact, $(T_0,r_+(T_0))=(\sqrt{2p/\pi},\sqrt{1/(8\pi p)})$ is exactly the meeting point of the $r_+$--$T$ curves of large and small black holes, as shown in Fig. \ref{f:rTSAdSwithout}. Because the black hole heat capacity satisfies $C_p=T\big(\f{\p S}{\p T}\big)_p=T\big(\f{\p S}{\p r_+}\big)_p\big(\f{\p r_+}{\p T}\big)_p$, we observe from Fig. \ref{f:rTSAdSwithout} that the large Schwarzschild--AdS black hole with positive $C_p$ is thermodynamically stable and can thus establish the equilibrium with the thermal AdS background, but a small one is on the contrary.

\begin{figure}[h]
\begin{center}
\includegraphics[width=0.45\textwidth]{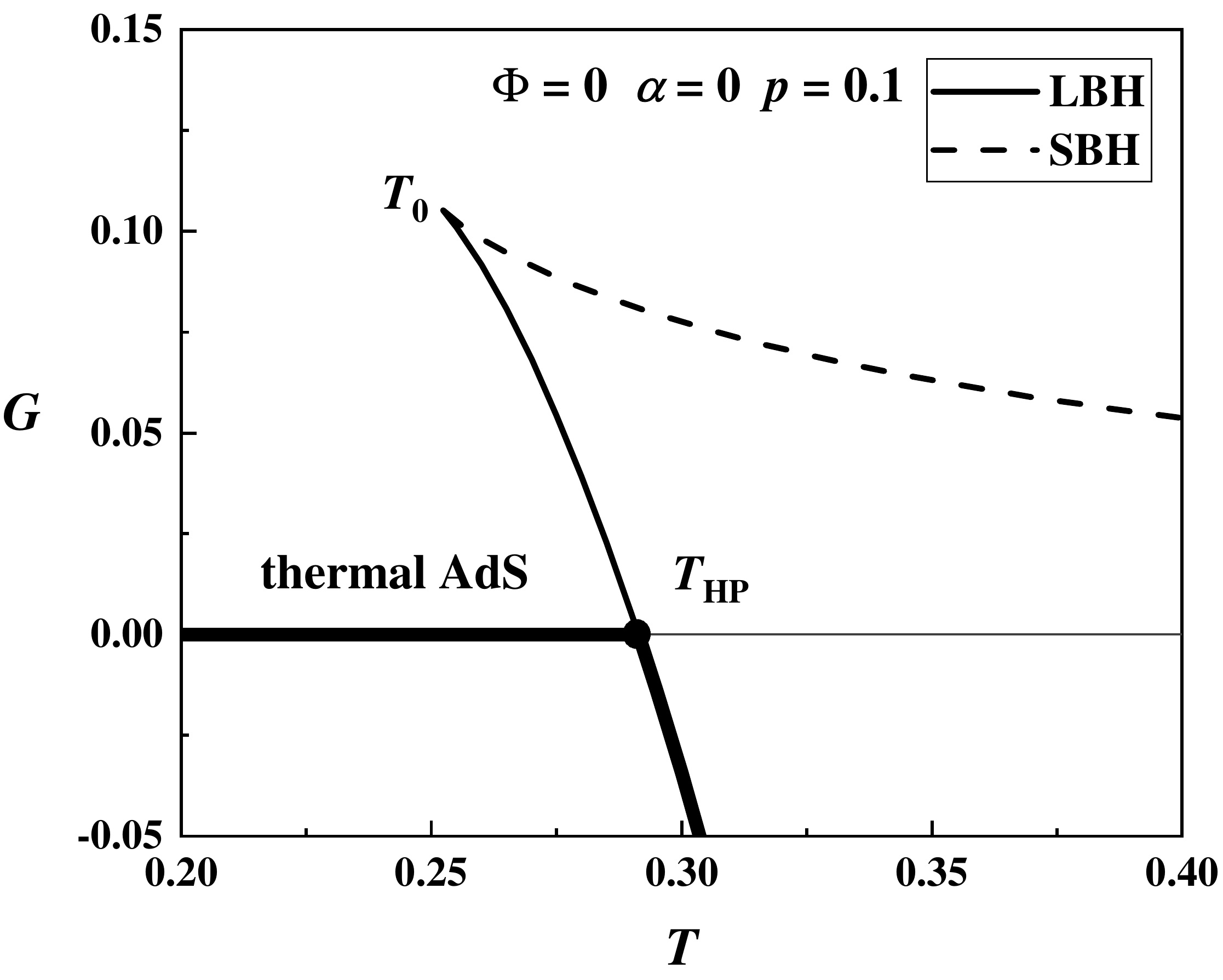}
\end{center}
\caption{The Gibbs free energies of large and small Schwarzschild--AdS black holes as a function of temperature, with a fixed pressure $p=0.1$. There is no HP phase transition for the small black hole, as its $G$--$T$ curve is always above the $T$-axis. At low or high temperatures, the thermal AdS phase or the large black hole phase is thermodynamically preferred respectively, with the global minimum of $G$ (thick black line), and their $G$--$T$ curves intersect at the HP temperature $T_{\rm HP}$, corresponding to a first-order phase transition.} \label{f:GTSAdSwithout}
\end{figure}

Next, substituting Eq. (\ref{rTSAdSwithout}) into Eq. (\ref{GSAdSwithout}), we obtain the Gibbs free energies of large and small Schwarzschild--AdS black holes,
\begin{flalign}
&G(T,p)=\frac{\sqrt{\pi T^2-p\pm T\sqrt{\pi^2 T^2-2\pi p}}}{24\sqrt{2\pi}p^2} \n\\
&\quad\quad\quad\quad\times\lt(4p- \pi T^2\mp T\sqrt{\pi^2 T^2-2\pi p}\rt). &\label{GTSAdSwithout}
\end{flalign}
The $G$--$T$ curves are shown in Fig. \ref{f:GTSAdSwithout}, and both curves decrease with temperature, meeting at $T_0$ with a cusp. For the unstable small black hole, its $G$--$T$ curve is concave and is always above the $T$-axis, without the HP phase transition. However, for the stable large black hole, its $G$--$T$ curve is convex and crosses the $T$-axis at the HP temperature $T_{\rm HP}$. Below or above $T_{\rm HP}$, the thermal AdS phase with vanishing $G$ or the black hole phase with negative $G$ is globally preferred respectively. At $T_{\rm HP}$, there is a discontinuity in the derivatives of the $G$--$T$ curves, indicating that the HP phase transition is of first-order.

Since the unstable small black holes cannot be in equilibrium with the thermal AdS background, we will not mention their HP phase transitions anymore.

\subsection{4EGB gravity} \label{sec:SAdSan0}

Now, we take into account the GB term and discuss its effects on the HP phase transitions in the 4EGB gravity. From Eqs. (\ref{T}) and (\ref{G}), the black hole temperature and Gibbs free energy are modified to
\begin{flalign}
%M&=\f{3r_+^2+3\alpha+8\pi pr_+^4}{6r_+}, \label{MSAdSwith}\\
&T=\f{r_+^2-\alpha+8\pi pr_+^4}{4\pi r_+^3}, &\label{TSAdSwith}\\
&G=\f{3r_+^2+3\alpha+8\pi pr_+^4}{6r_+} \n\\
&\quad\quad -\f{r_+^2-\alpha+8\pi p r_+^4}{4 r_+(r_+^2+2\alpha)}\lt(r_+^2+4\alpha\ln\f{r_+}{\sqrt\alpha}\rt). &\label{GSAdSwith}
\end{flalign}

Before the relevant discussions, we should first point out an important difference between the Einstein and 4EGB gravities. In the former, from Eq. (\ref{TpSAdSwithout}), there is no lower or upper bound of $p$, and the HP phase transition can occur at all pressures. However, in the latter, $p$ has both lower and upper bounds, making the analyses more complicated, so we must determine the range of $p$ in advance. This range comes from three aspects. First, we expect the HP phase transition above the critical pressure of the Schwarzschild--AdS black hole, otherwise a swallowtail behavior will appear in the $G$--$T$ curve, and the black hole will undergo a van der Waals-like phase transition, which is beyond the scope of our present work. This requirement will set a lower bound of $p$. Second, we will see immediately that, when $\alpha\neq 0$, the minimal black hole temperature can reach zero, so the minimal black hole entropy should be evaluated as $S(0,p,\alpha)$. The positivity of $S(0,p,\alpha)$ will set an upper bound of $p$. Third, since the Gibbs free energy decreases with temperature, if the HP phase transition occurs, $G(0,p,\alpha)$ should also be positive, but we will find that this condition is satisfied automatically. Altogether, $p$ has both lower and upper bounds simultaneously.

First, the critical pressure $p_{\rm c}$ can be fixed by the equation of state in Eq. (\ref{eos}),
\begin{flalign}
&\lt(\f{\p p}{\p r_+}\rt)_{T,\alpha}=\lt(\f{\p^2 p}{\p r_+^2}\rt)_{T,\alpha}=0, &\n
\end{flalign}
and these conditions set the lower bound of $p$,
%\footnote{Here, we should mention that, in four-dimensional Einstein gravity, the Schwarzschild--AdS black holes have no critical phenomena, but only the RN--AdS black holes have, as all the critical temperature, volume, and pressure are the function of black hole charge. However, in the 4EGB gravity, even the Schwarzschild--AdS black holes can have critical phenomena, because now all the critical values are the function of $\alpha$, even if the black holes are neutral. This is also an interesting difference between the Einstein and 4EGB gravities.}
\begin{flalign}
&p>p_{\rm c}=\frac{15-8\sqrt{3}}{288\pi\alpha}\approx\f{0.00126}{\alpha}. &\n
\end{flalign}
%\begin{flalign}
%r_{\rm c}&=\sqrt{6\alpha+4\sqrt{3}\alpha},\n\\
%T_{\rm c}&=\frac{\sqrt{2\sqrt{3}-3}}{6\pi\sqrt{2\alpha}},\n\\
%\end{flalign}
Second, when $T=0$, from Eq. (\ref{TSAdSwith}), we obtain
\begin{flalign}
&r_+=\sqrt{\f{\sqrt{1+32\pi\alpha p}-1}{16\pi p}}. &\label{rrr}
\end{flalign}
Substituting Eq. (\ref{rrr}) into Eq. (\ref{S}), we need
\begin{flalign}
&S(0,p,\alpha)=\f{\sqrt{1+32\pi\alpha p}-1}{16p}\n\\
&\qquad\qquad\quad+2\pi\alpha\ln\f{\sqrt{1+32\pi\alpha p}-1}{16\pi\alpha p}>0. &\n
\end{flalign}
This inequality can only be solved numerically,
\begin{flalign}
&p<\f{0.0238}{\alpha}. &\n
\end{flalign}
Third, substituting Eq. (\ref{rrr}) into Eq. (\ref{GSAdSwith}), we easily find
\begin{flalign}
&G(0,p,\alpha)=\f{\sqrt{1+32\pi\alpha p}-1+32\pi\alpha p}{12\sqrt{\pi p(\sqrt{1+32\pi\alpha p}-1)}}>0, &\label{jieju} %=\f{r_+^2+2\alpha}{3r_+}
\end{flalign}
so this inequality does not provide any further constraint. In all, the range of $p$ can be summarized as
\begin{flalign}
&0.00126<\alpha p<0.0238. &\label{range}
\end{flalign}
In the following, we restrict our discussions within this range, in which the black holes can have and only have the HP phase transitions. %Beyond this range, the black holes either have no HP phase transition, or have the van der Waals-like phase transition.

\begin{figure}[h]
\begin{center}
\includegraphics[width=0.45\textwidth]{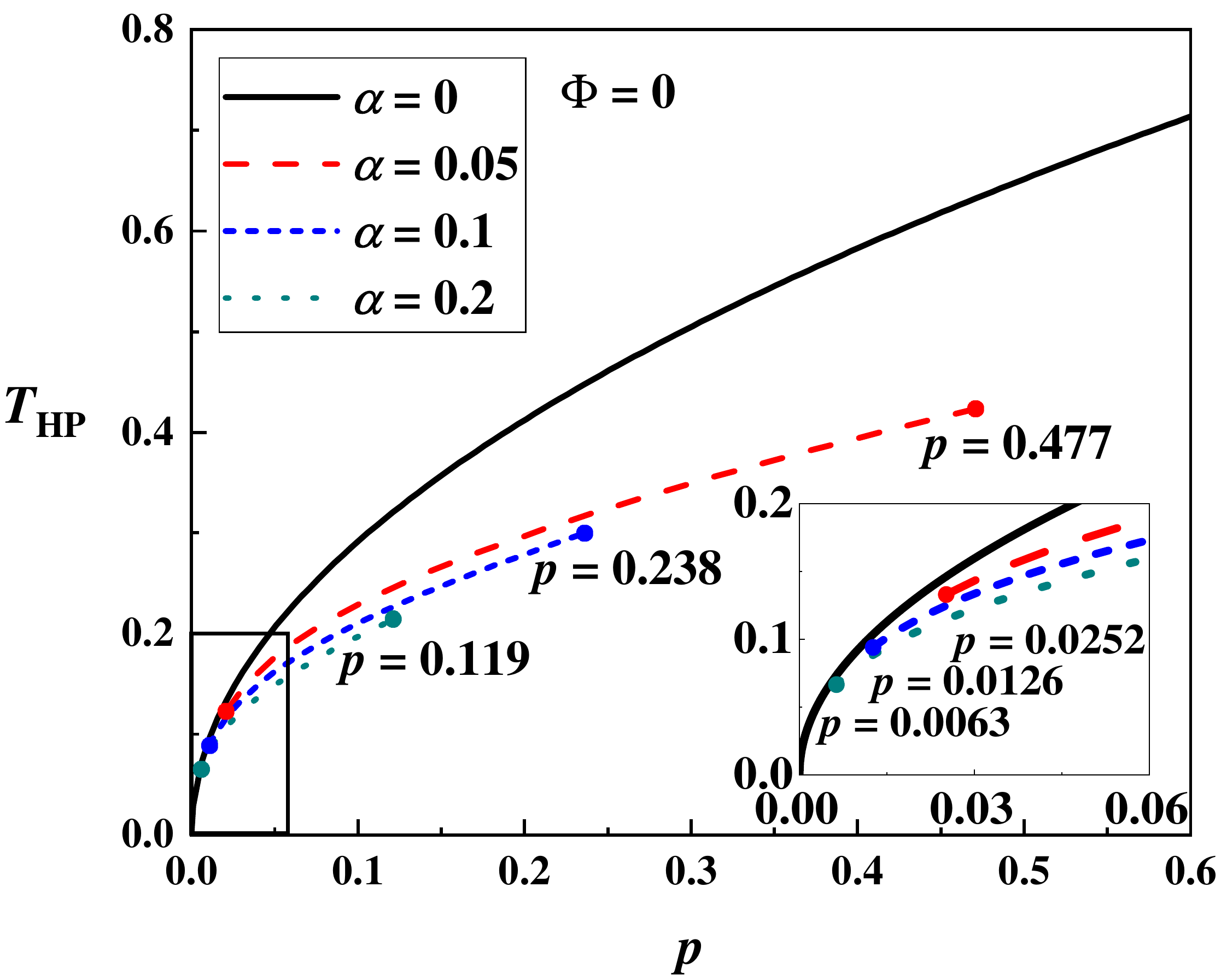}
\end{center}
\caption{The HP temperature of the Schwarzschild--AdS black holes as a function of pressure. The solid line is the coexistence line in the Einstein gravity. In the 4EGB gravity, there are both lower and upper bounds of $p$ as shown in Eq. (\ref{range}), with the detailed values in the figure, corresponding to the terminal points in the coexistence lines. At a fixed pressure $p$, $T_{\rm HP}$ decreases with $\alpha$.} \label{f:TpSAdSwith}
\end{figure}

Now, we return to the HP phase transitions in the 4EGB gravity. By the same procedure in Sect. \ref{sec:SAdSa0}, the $T_{\rm HP}$--$p$ curves can be numerically plotted in Fig. \ref{f:TpSAdSwith}. We find that, for any non-vanishing $\alpha$, the HP phase transitions can only occur in the range in Eq. (\ref{range}), and at a given pressure, the HP temperature $T_{\rm HP}$ decreases with $\alpha$.

\begin{figure}[h]
\begin{center}
\includegraphics[width=0.45\textwidth]{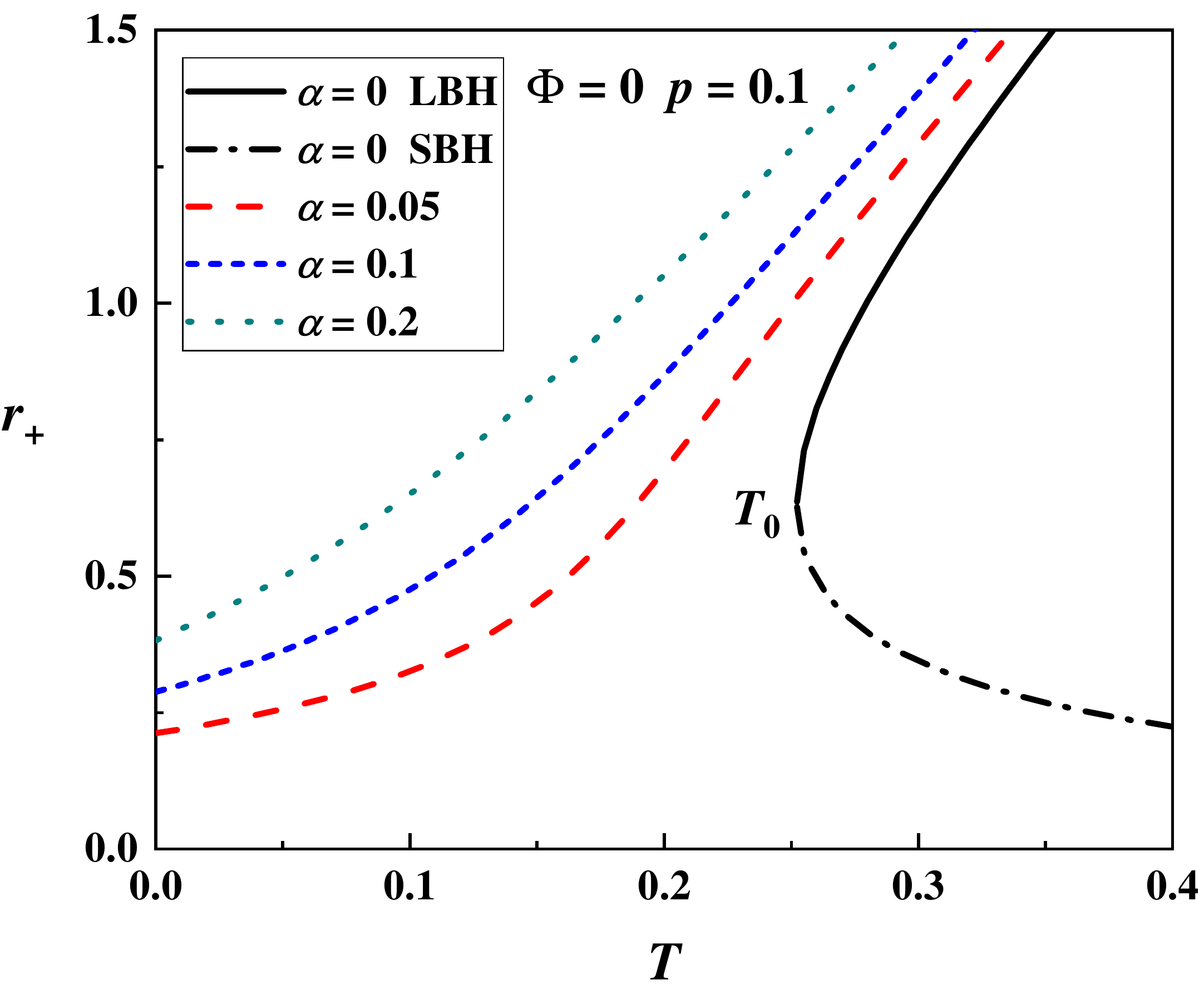}
\end{center}
\caption{The event horizon radii of the Schwarzschild--AdS black holes as a function of temperature in the 4EGB gravity, with a fixed pressure $p=0.1$ and different values of $\alpha$. Unlike the Einstein gravity, the minimal black hole temperature is no longer $T_0=\sqrt{2p/\pi}$, but can reach zero.} \label{f:rTSAdSwith}
\end{figure}

Next, from Eq. (\ref{TSAdSwith}), the $r_+$--$T$ curves are shown in Fig. \ref{f:rTSAdSwith}. We see that, totally different from Fig. \ref{f:rTSAdSwithout}, the minimal temperatures of the Schwarzschild--AdS black holes can approach zero now. This is the distinct difference between the Einstein and 4EGB gravities, and the reason is that there is a negative term $-\alpha$ in the numerator in Eq. (\ref{TSAdSwith}), but not in Eq. (\ref{TSAdSwithout}).

The difference of the minimal black hole temperature causes the $G$--$T$ curves in the 4EGB gravity obviously different from those in the Einstein gravity in Fig. \ref{f:GTSAdSwithout}. Here, all the $G$--$T$ curves set out from the $G$-axis with $T=0$, instead of $T_0$. From Eq. (\ref{jieju}), the intercept on the $G$-axis $G(0,p,\alpha)$ increases with $\alpha$. Meanwhile, the intercept on the $T$-axis (i.e., the HP temperature $T_{\rm HP}$) decreases with $\alpha$, consistent with Fig. \ref{f:TpSAdSwith}.
\begin{figure}[h]
\begin{center}
\includegraphics[width=0.45\textwidth]{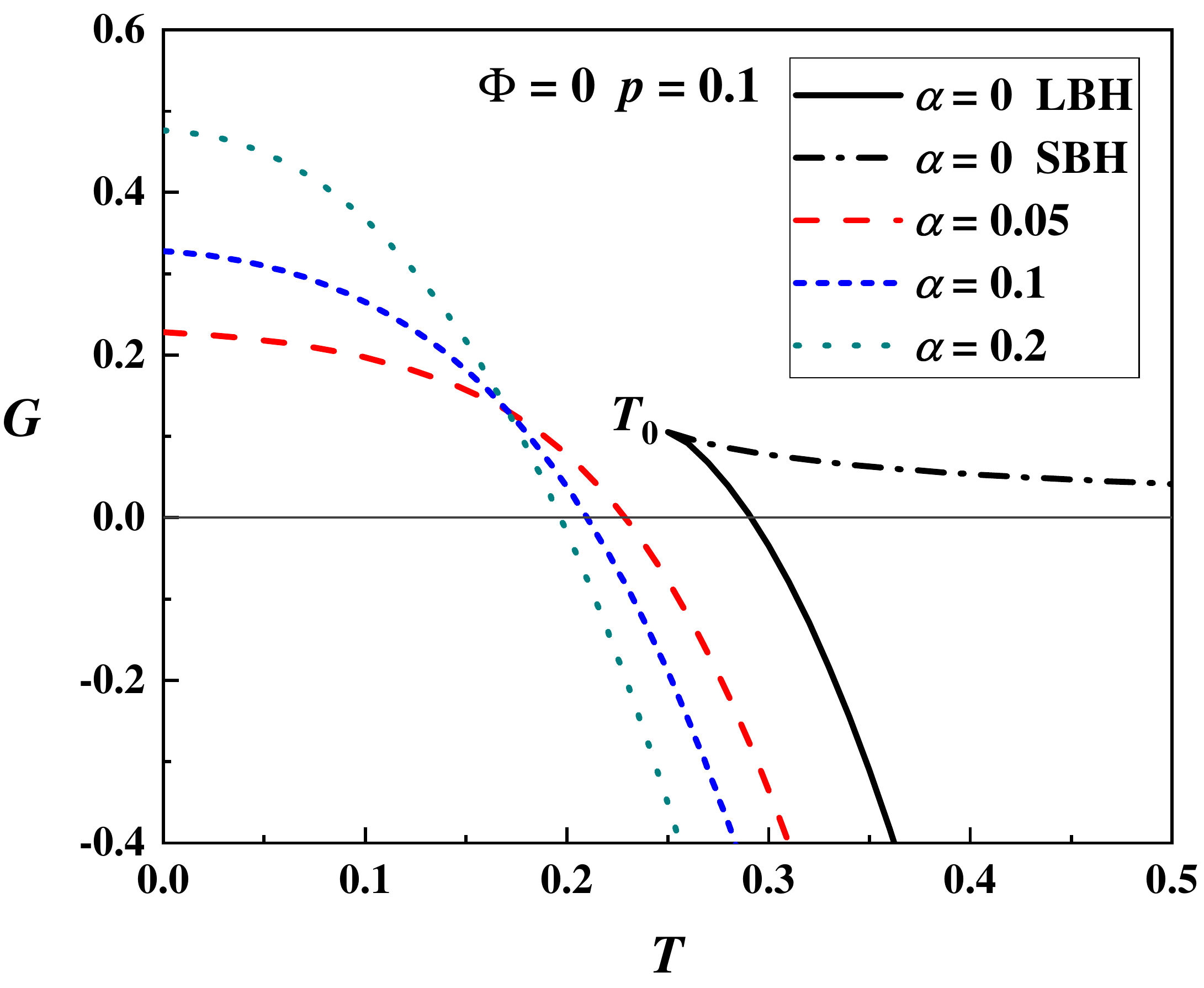}
\end{center}
\caption{The Gibbs free energies of the large Schwarzschild--AdS black holes as a function of temperature in the 4EGB gravity, with a fixed pressure $p=0.1$ and different values of $\alpha$. All the $G$--$T$ curves start from the $G$-axis, with the vanishing black hole temperature. Moreover, from the intersection points of the $G$--$T$ curves across the $T$-axis, the HP temperature $T_{\rm HP}$ decreases with $\alpha$. The $G$--$T$ curves in the Einstein gravity are also shown for comparison.} \label{f:GTSAdSwith}
\end{figure}

\section{HP phase transitions of the RN--AdS black holes} \label{sec:HPRNAdS}

In this section, we continue our discussions on the HP phase transitions of the RN--AdS black holes in the Einstein and 4EGB gravities. Before detailed calculations, one important issue should be stressed. On account of the conservation of charge and the neutrality of thermal AdS background, the HP phase transitions of the charged AdS black holes must be considered in the grand canonical ensemble with fixed electric potential.

\subsection{Einstein gravity} \label{sec:RNAdSa0}

The discussions in the Einstein gravity are in parallel with Sect. \ref{sec:SAdSa0}. The results were also given in Ref. \cite{HPwo}, and we only list them briefly. The metric of the RN--AdS black hole is shown in Eq. (\ref{RN}), from which we can determine the range of black hole charge as $Q<M$ and the range of electric potential as $\Phi<1$.

From Eqs. (\ref{T}) and (\ref{G}), the RN--AdS black hole temperature and Gibbs free energy read
\begin{flalign}
%M&=\f{3(1+\Phi^2)r_++8\pi pr_+^3}{6}, \label{MRNAdSwithout}\\
&T=\f{1-\Phi^2+8\pi p r_+^2}{4\pi r_+}, &\label{TRNAdSwithout}\\
&G=M-TS-\Phi Q=\f{3(1-\Phi^2)r_+ -8\pi pr_+^3}{12}.&\label{GRNAdSwithout}
\end{flalign}
From Eq. (\ref{panju}), at the HP phase transition point, $r_+=\sqrt{3(1-\Phi^2)/(8\pi p)}$, and the HP temperature is
\begin{flalign}
&T_{\rm HP}=\sqrt{\f{8p}{3\pi}(1-\Phi^2)}. &\label{TpRNAdSwithout}
\end{flalign}
Again, there is no bound of $p$, so the HP phase transition can occur at all pressures, and at a given pressure, $T_{\rm HP}$ decreases with $\Phi$.

\begin{figure}[h]
\begin{center}
\includegraphics[width=0.45\textwidth]{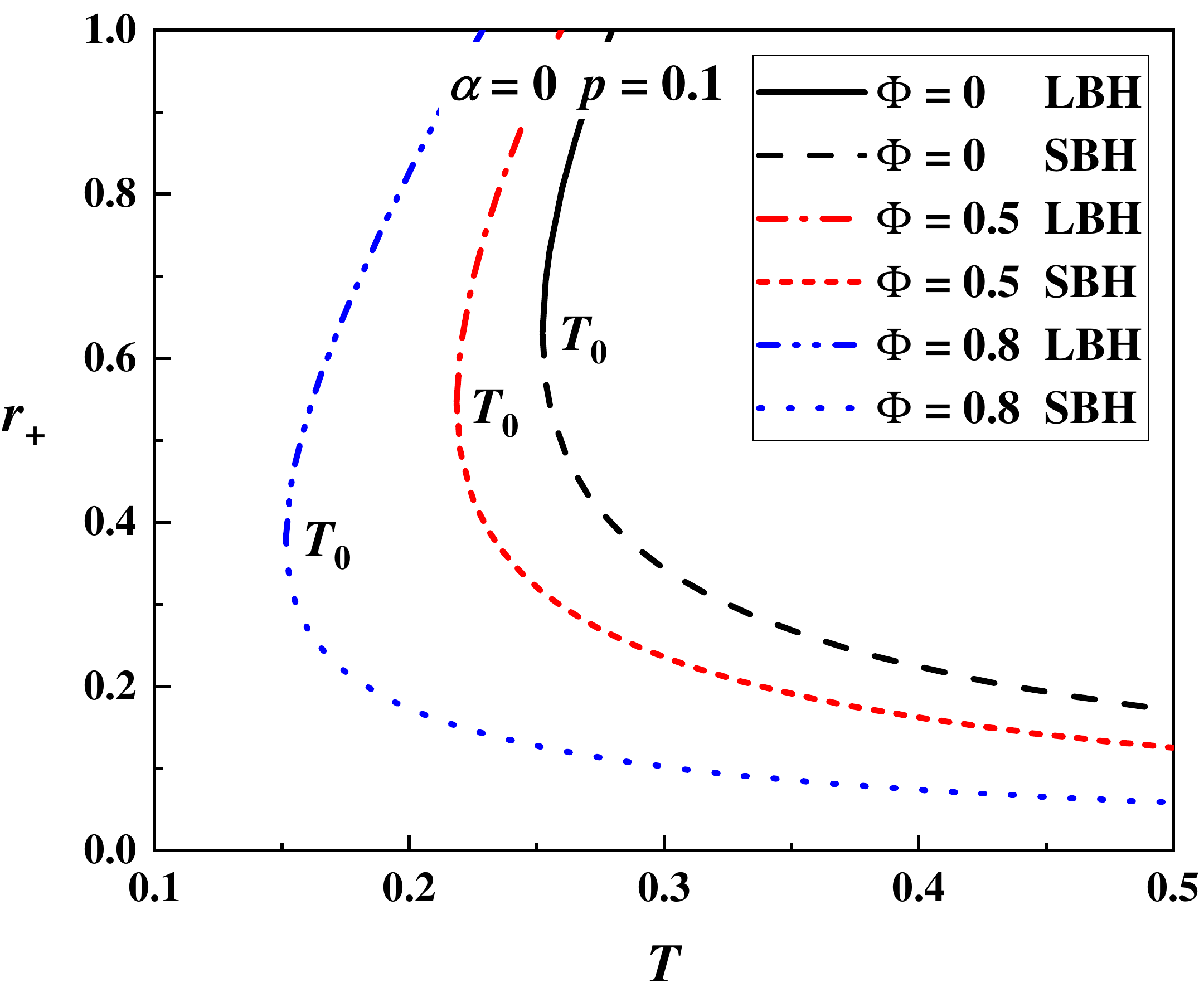}
\end{center}
\caption{The event horizon radii of the RN--AdS black holes as a function of temperature, with a fixed pressure $p=0.1$ and different values of $\Phi$. Similar to the Schwarzschild--AdS case, the RN--AdS black hole also has a minimal temperature, but modified by the electric potential as $T_0=\sqrt{2p(1-\Phi^2)/\pi}$.} \label{f:rTRNAdSwithout}
\end{figure}

Next, from Eq. (\ref{TRNAdSwithout}), we have
\begin{flalign}
&r_+(T,p)=\f{1}{4\pi p}\lt[\pi T \pm \sqrt{\pi^2 T^2-2\pi p(1-\Phi^2)}\rt], &\label{rTRNAdSwithout}
\end{flalign}
and the minimal black hole temperature is
\begin{flalign}
&T_0=\sqrt{\f{2p}{\pi}(1-\Phi^2)}. &\n % \label{t0}
\end{flalign}
The $r_+$--$T$ curves are plotted in Fig. \ref{f:rTRNAdSwithout}, which are similar to those in Fig. \ref{f:rTSAdSwithout}.

\begin{figure}[h]
\begin{center}
\includegraphics[width=0.45\textwidth]{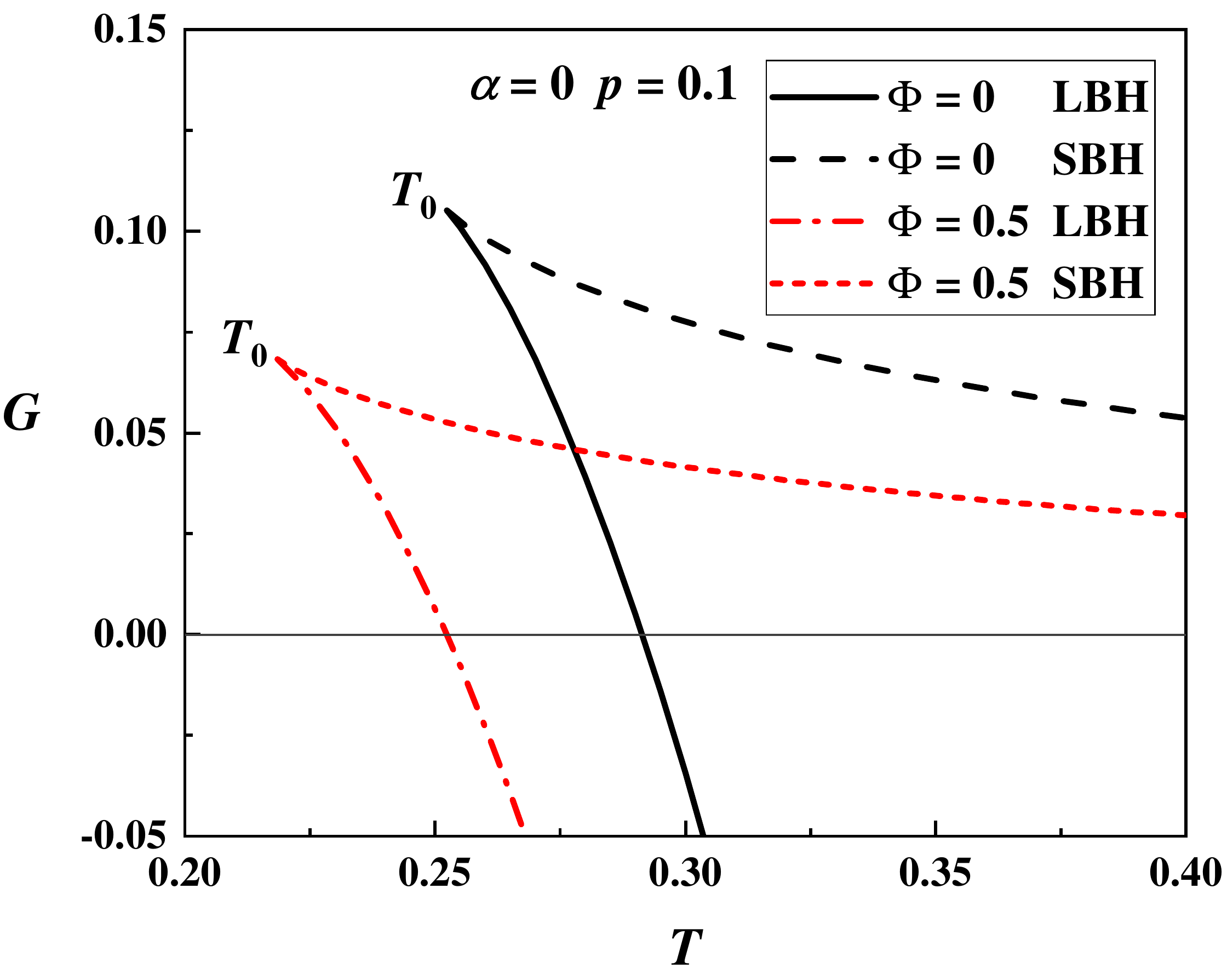}
\end{center}
\caption{The Gibbs free energies of large and small RN--AdS black holes as a function of temperature, with a fixed pressure $p=0.1$ and different values of $\Phi$. The HP temperature $T_{\rm HP}$ decreases with the electric potential $\Phi$, as shown in Eq. (\ref{TpRNAdSwithout}).} \label{f:GTRNAdSwithout}
\end{figure}

Substituting Eq. (\ref{rTRNAdSwithout}) into Eq. (\ref{GRNAdSwithout}), the Gibbs free energies of the large and small RN--AdS black holes are
%\begin{flalign}
%&\quad G(T,p,\Phi) \n\\
%&G(T,p,\Phi)=\frac{\sqrt{\pi T^2-p(1-\Phi^2)\pm T\sqrt{\pi^2 T^2-2\pi p(1-\Phi^2)}}}{24\sqrt{2\pi}p^2} \n\\
%&\quad\quad\quad\quad\quad \times\lt[4p(1-\Phi^2)- \pi T^2\mp T\sqrt{\pi^2 T^2-2\pi p(1-\Phi^2)}\rt].&\n
%\end{flalign}
\begin{flalign}
%&\quad G(T,p,\Phi) \n\\
&G(T,p,\Phi) \n\\
&=\frac{\sqrt{\pi T^2-p(1-\Phi^2)\pm T\sqrt{\pi^2 T^2-2\pi p(1-\Phi^2)}}}{24\sqrt{2\pi}p^2} \n\\
&\quad \times\lt[4p(1-\Phi^2)- \pi T^2\mp T\sqrt{\pi^2 T^2-2\pi p(1-\Phi^2)}\rt].&\n
\end{flalign}
The $G$--$T$ curves are shown in Fig. \ref{f:GTRNAdSwithout}. We see that the RN--AdS black hole still has a minimal temperature $T_0$, and the HP temperature $T_{\rm HP}$ decreases with $\Phi$, consistent with Eq. (\ref{TpRNAdSwithout}).

\subsection{4EGB gravity} \label{sec:RNAdSan0}

Finally, we move on to the most general case, the HP phase transitions of the RN--AdS black holes in the 4EGB gravity. From Eqs. (\ref{T}) and (\ref{G}), we have
\begin{flalign}
%M&=\f{3(1+\Phi^2)r_+^2+3\alpha+8\pi pr_+^4}{6r_+}, \label{MRNAdSwith}\\
&T=\f{(1-\Phi^2)r_+^2-\alpha+8\pi p r_+^4}{4\pi r_+(r_+^2+2\alpha)}, &\label{TRNAdSwith}\\
&G=\f{3(1-\Phi^2)r_+^2+3\alpha+8\pi pr_+^4}{6r_+} \n\\
&\qquad -\f{(1-\Phi^2)r_+^2-\alpha+8\pi p r_+^4}{4 r_+(r_+^2+2\alpha)}\lt(r_+^2+4\alpha\ln\f{r_+}{\sqrt\alpha}\rt). &\label{GRNAdSwith}
\end{flalign}

\begin{figure}[h]
\begin{center}
\includegraphics[width=0.45\textwidth]{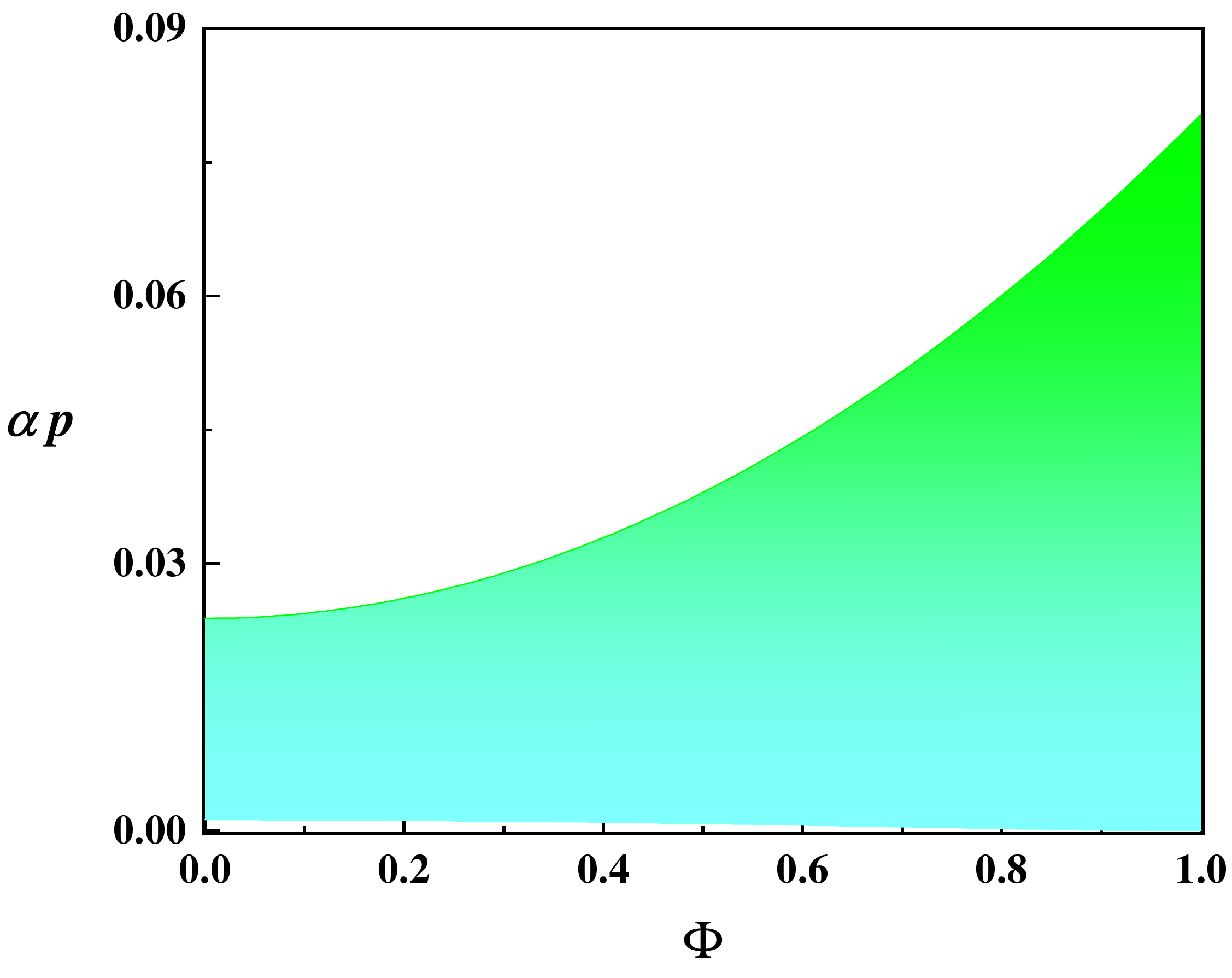}
\end{center}
\caption{The range of pressure of the RN--AdS black holes in the HP phase transitions in the 4EGB gravity. To be convenient and consistent with Eq. (\ref{range}), we show the relation of $\alpha p$ and $\Phi$, and the HP phase transitions can occur only in the shadowed area. When $\Phi=0$, the intercepts 0.00126 and 0.0238 recover the range in the Schwarzschild--AdS case in Eq. (\ref{range}). With $\Phi$ increasing, the range of $p$ allowed for the HP phase transition becomes larger.} \label{f:apPhi}
\end{figure}

Again, we need to determine the range of pressure $p$ in advance. Following the discussions in Sect. \ref{sec:SAdSan0}, first, $p$ should be greater than its lower bound (i.e., critical pressure $p_{\rm c}$), such that there is only HP phase transition, but no van der Waals-like one. Second, from $S(0,p,\Phi,\alpha)>0$, we can fix the upper bound of $p$. Third, the condition $G(0,p,\Phi,\alpha)>0$ still does not provide further constraint. Now, the range of $p$ depends not only on the GB coupling constant $\alpha$, but also on the electric potential $\Phi$, as numerically shown in Fig. \ref{f:apPhi}. We find that, with $\Phi$ increasing, the range of $p$ becomes larger, so the HP phase transitions can occur in a larger range of pressure.
%\begin{flalign}
%r_{\rm c}&=\sqrt{\frac{3(2-\Phi^2)\alpha+\sqrt{3}\sqrt{3\Phi^4-16\Phi^2+16}\alpha}{1-\Phi^2}},\label{123}\\
%T_{\rm c}&=\frac{\sqrt{3(2-\Phi^2)+\sqrt{3}\sqrt{3\Phi^4-16\Phi^2+16}}}{48\pi\sqrt{\alpha(1-\Phi^2)}}\n\\
%&\quad\times\lt[(8-5\Phi^2)-\sqrt{3}\sqrt{3\Phi^4-16\Phi^2+16}\rt],\label{231}\\
%p_{\rm c}&=\frac{(1-\Phi^2)^2\lt[3(3-\Phi^2)+\sqrt{3}\sqrt{3\Phi^4-16\Phi^2+16}\rt]} %{24\pi\alpha\lt[3(2-\Phi^2)+\sqrt{3}\sqrt{3\Phi^4-16\Phi^2+16}\rt]^2}, \label{312}
%\end{flalign}

\begin{figure}[h]
\begin{center}
\includegraphics[width=0.45\textwidth]{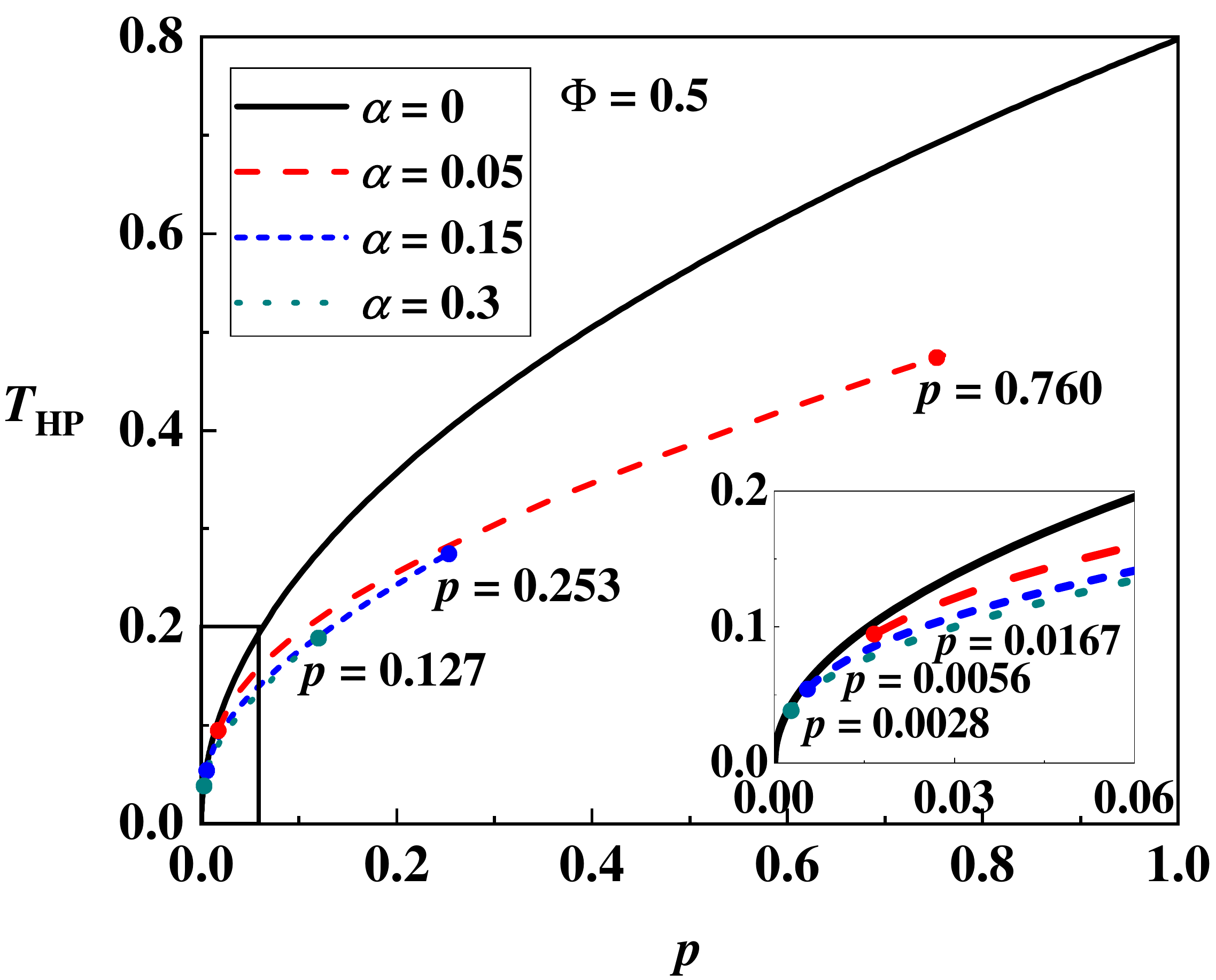}
\end{center}
\caption{The HP temperature of the RN--AdS black holes as a function of pressure. The solid line is the coexistence line in the Einstein gravity. In the 4EGB gravity, pressure $p$ has both lower and upper bounds, with the detailed values in the figure. At a fixed pressure $p$, $T_{\rm HP}$ decreases with both $\alpha$ and $\Phi$.} \label{f:TpRNAdSwith}
\end{figure}

\begin{figure}[h]
\begin{center}
\includegraphics[width=0.45\textwidth]{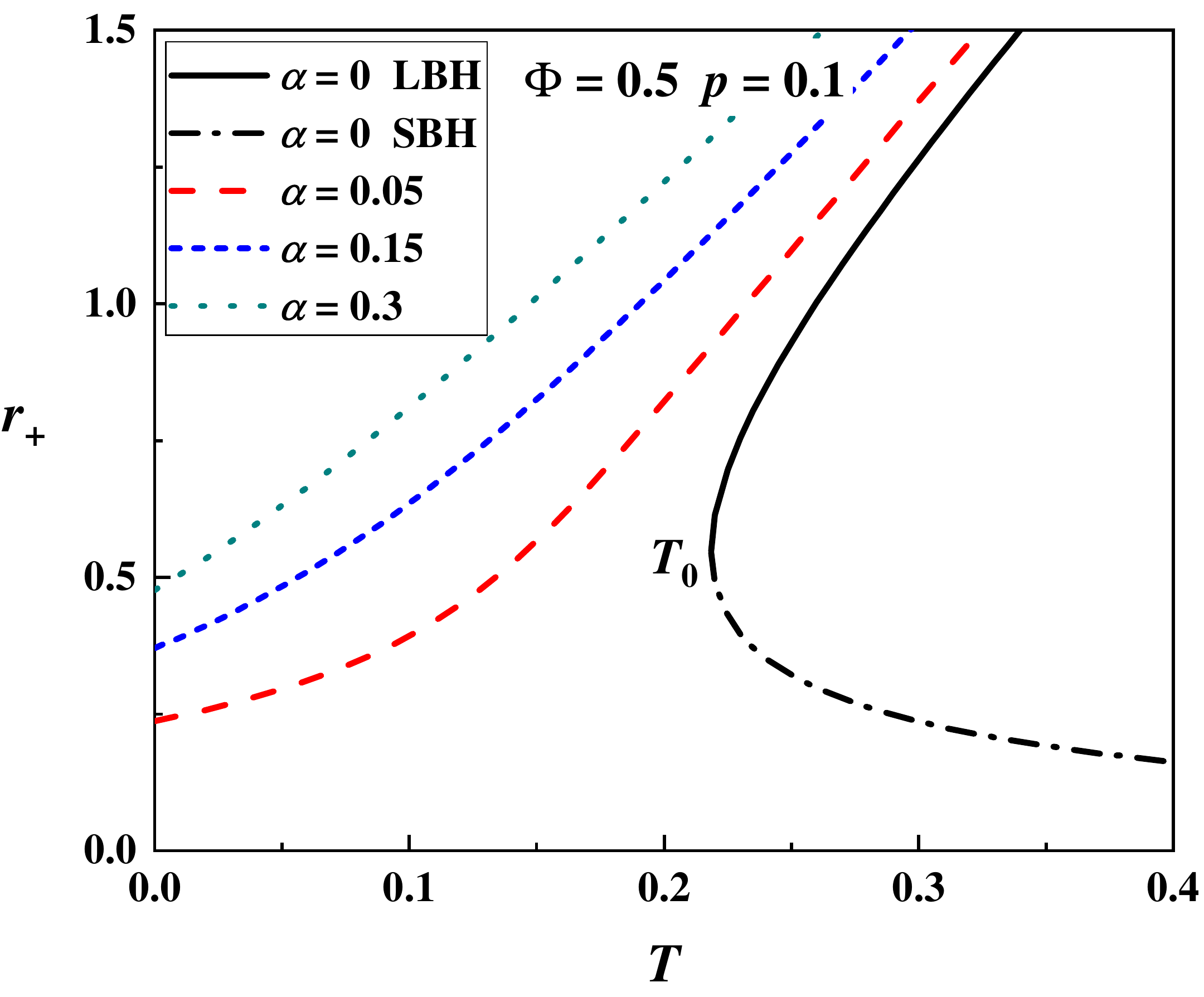}
\end{center}
\caption{The event horizon radii of the RN--AdS black holes as a function of temperature in the 4EGB gravity, with a fixed pressure $p=0.1$, fixed electric potential $\Phi=0.5$, and different values of $\alpha$. The minimal black hole temperature is zero.} \label{f:rTRNAdSwith}
\end{figure}

\begin{figure}[!h]
\begin{center}
\includegraphics[width=0.45\textwidth]{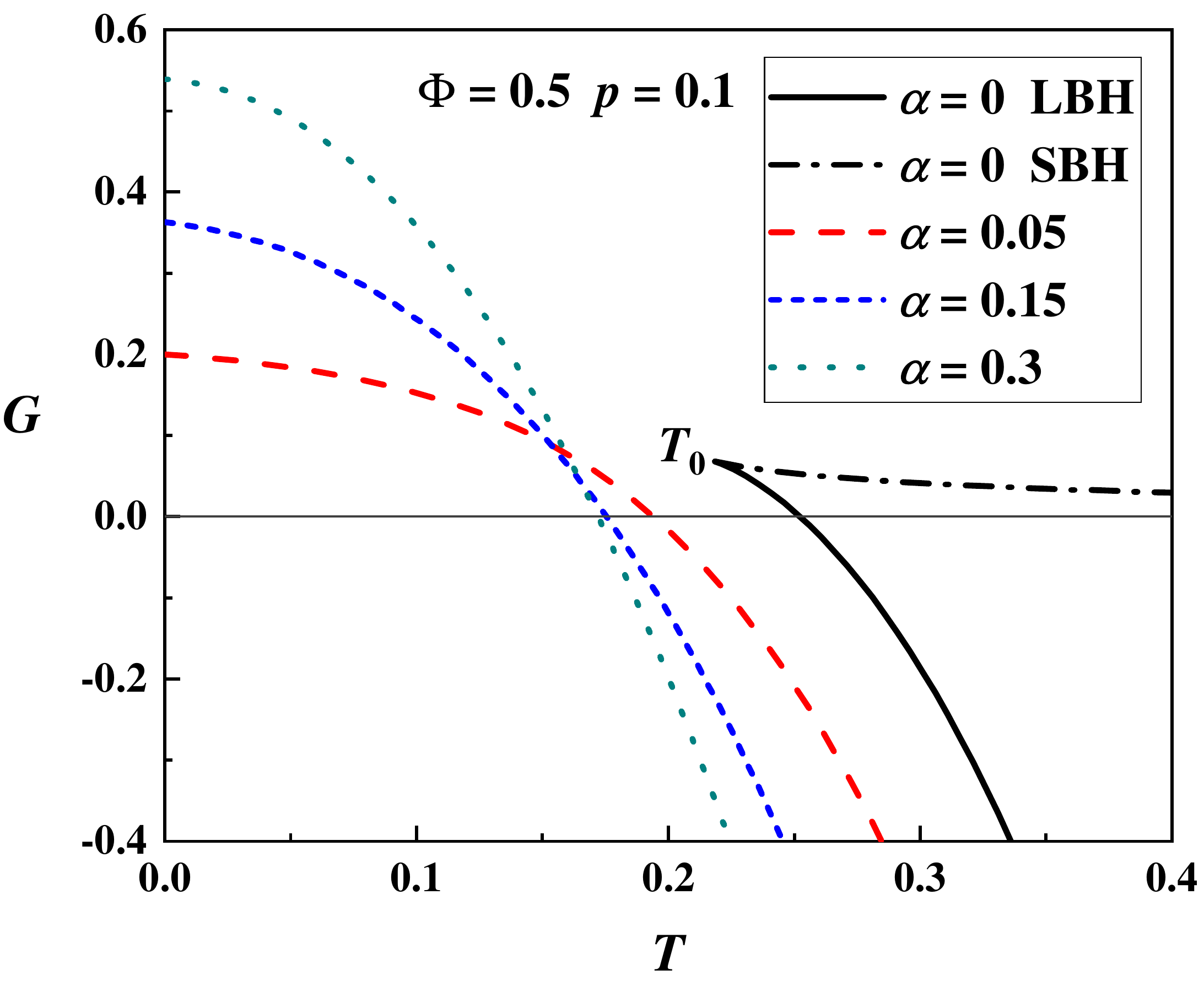}
\end{center}
\caption{The Gibbs free energies of the large RN--AdS black holes as a function of temperature in the 4EGB gravity, with a fixed pressure $p=0.1$, fixed electric potential $\Phi=0.5$, and different values of $\alpha$. The HP temperature $T_{\rm HP}$ decreases with both $\alpha$ and $\Phi$. The $G$--$T$ curves in the Einstein gravity are also shown for comparison.} \label{f:GTRNAdSwith}
\end{figure}

With the above preparations, the investigations of the HP phase transitions of the RN--AdS black holes in the 4EGB gravity are straightforward. In the following, we choose the electric potential $\Phi=0.5$ and plot the $T_{\rm HP}$--$p$, $r_+$--$T$, and $G$--$T$ curves in Figs. (\ref{f:TpRNAdSwith})--(\ref{f:GTRNAdSwith}) respectively.

From Figs. (\ref{f:TpRNAdSwith})--(\ref{f:GTRNAdSwith}), we observe that the HP temperature $T_{\rm HP}$ decreases with both $\alpha$ and $\Phi$, and the minimal black hole temperature can reach zero. These general features are similar to those in the Schwarzschild--AdS case in the 4EGB gravity, only with the detailed values of $T_{\rm HP}$ and the range of $p$ modified by $\Phi$.

Till now, we complete the discussions of the HP phase transitions of the RN--AdS black holes in the 4EGB gravity. We should emphasize that the similarities to the Schwarzschild--AdS case are based upon the fact that we utilize the grand canonical ensemble with fixed electric potential. Otherwise, if we chose the canonical ensemble with fixed charge, the situations would become different, but such choice is not quite physically meaningful, see the Appendix in Ref. \cite{HPwo}.

Last, we briefly comment the HP phase transitions of the rotating AdS black holes in the 4EGB gravity. Unfortunately, the rotating black hole metric solution in the 4EGB gravity is still unavailable. Although there have been some attempts via the Newman--Janis algorithm \cite{Wei:2020ght, Kumar:2020owy}, the effective rotating solution is not obtained by solving the field equations. Meanwhile, from the results in Sects. \ref{sec:HPSAdS} and \ref{sec:HPRNAdS} and also from our experience in Ref. \cite{HPwo}, except the mathematical complexities caused by rotation, we do not expect more new physical results.

\section{Conclusion} \label{sec:con}

Black hole thermodynamics in the extended phase space has received intensive research interests, as there are much richer thermodynamic phenomena due to the introduction of the effective black hole volume and pressure. For example, the HP phase transitions in the extended phase space are widely studied within and beyond the Einstein gravity.

The gravity theory with the GB term is the minimal extension of the Einstein gravity. In four-dimensional GB gravity, the GB term is a topological invariant and is irrelevant to gravitational dynamics. However, it changes black hole thermodynamics via correcting the black hole entropy and thus influences the HP phase transition. These were the results in our previous work in Ref. \cite{HPwo}.

Recently, a novel 4EGB gravity theory was realized by rescaling the GB coupling constant $\alpha\to\alpha/(d-4)$ in $d$ dimensions and redefining the four-dimensional gravity in the limit $d \to 4$ \cite{4EGB}. In such a way, the GB term has nontrivial dynamical effects even in four dimensions, and both the black hole mass and entropy are altered accordingly. As a result, the HP phase transition will exhibit new characters in the 4EGB gravity, and this is the research topic of our present work.

In this paper, we study the HP phase transitions of the Schwarzschild--AdS and RN--AdS black holes in the extended phase space. Comparing the results in the Einstein and 4EGB gravities, we find similarities and dissimilarities at the same time. The similarities are that the HP temperature $T_{\rm HP}$ increases with pressure $p$, and the Gibbs free energy $G$ decreases with temperature $T$. However, in the Einstein gravity, $p$ has no bound, and the black hole has a positive minimal temperature $T_0$, but in the 4EGB gravity, $p$ has both lower and upper bounds, and minimal black hole temperature is zero. These differences induce the dissimilarities in the $T_{\rm HP}$--$p$ and $G$--$T$ curves, and $T_{\rm HP}$ is found to decrease with the GB coupling constant $\alpha$ and the electric potential $\Phi$.

Furthermore, it is very natural and interesting to generally compare the HP phase transitions in the Einstein, GB, and 4EGB gravities together. In the GB gravity, the GB term only changes the black hole entropy, but not mass. Moreover, $p$ has only an upper bound, and the minimal black hole temperature is still $T_0$, but not zero. Therefore, we can conclude that the HP phase transition behaviors in the GB gravity lie exactly between the Einstein and 4EGB gravities.

In summary, the Einstein, GB, and 4EGB gravities form a series, in which the influences from the GB term become greater and greater. These influences result in the tighter and tighter bounds of pressure, within which the HP phase transitions can occur. Altogether, we wish to provide a general picture of the HP phase transitions with GB term to the most extent.

\vskip .5cm
%\begin{acknowledgements}
We are very grateful to Zhao-Hui Chen, Yu-Xiao Liu, Shao-Wen Wei, and Tao Zhu for fruitful discussions. This work is supported by the Fundamental Research Funds for the Central Universities of China (No. N170504015).
%\end{acknowledgements}

\begin{appendix}

\section{Black hole entropy in the 4EGB gravity via the Iyer--Wald formula}

Here, we explicitly show that the result of the black hole entropy in the 4EGB gravity in Eq. (\ref{S}) can also be obtained via the Iyer--Wald formula \cite{Wald}.

We start from the action of the 4EGB gravity,
\begin{flalign}
&S_4=\f{1}{16\pi}\int\dd^4x\,\sqrt{-g}\big\{R-2\Lambda+\alpha[\phi{\cal G}+4G^{\mu\nu}\p_\mu\phi\p_\nu\phi &\n\\
&\qquad-4(\p\phi)^2\p^2\phi+2((\p\phi)^2)^2]\big\}. &\label{1}
%&\lt. \quad -2\lambda Re^{-2\phi}-12\lambda(\p\phi)^2e^{-2\phi}-6\lambda^2e^{-4\phi}
\end{flalign}
%and this action is a special scalar--tensor theory that belongs to the family of the Horndeski gravity. Here, we clearly observe the non-minimal coupling of the scalar field $\phi$ to the GB term ${\cal G}$ and also the nonlinear terms for $\phi$.
%From the action in Eq. (\ref{1}), the four-dimensional static and spherically symmetric black hole solutions have been obtained in Eq. (3.6) of Ref. \cite{Hennigar:2020lsl} and in Eq. (20) of Ref. \cite{lue}. These solutions are precisely the same as those obtained from the $d$-dimensional action as
%\begin{flalign}
%&\f{1}{16\pi G_d}\int\dd^d x\,{\sqrt{-g}}\lt(R-2\Lambda+\f{\alpha}{d-4}{\cal G}\rt), &\label{2}
%\end{flalign}
%and then taking the limit $d\to 4$. The same remains true in the presence of electromagnetic field. For the more complicated black holes without spherical horizon topology (e.g., the Taub--NUT black holes discussed in Ref. \cite{Hennigar:2020lsl}, the black hole solutions obtained from Eqs. (\ref{1}) and (\ref{2}) do not necessarily coincide. Nevertheless, in our present work, as we focus on the spherically symmetric black holes (i.e., the Schwarzschild--AdS and RN--AdS black holes), the black hole solutions are the same from Eqs. (\ref{1}) and (\ref{2}). So are their entropies.
As for the Iyer--Wald formula, the static and spherically symmetric black hole metric is
\begin{flalign}
&\dd s^2_4=-f(r)\,\dd t^2+\f{\dd r^2}{f(r)}+r^2\,\dd\Omega^2. &\label{3}
\end{flalign}
Substituting Eq. (\ref{3}) into Eq. (\ref{1}), we obtain an effective Lagrangian in terms of $f(r)$ and $\phi(r)$,
\begin{flalign}
& 2(1-\Lambda r^2-f-rf')+\f 23\alpha\phi'[3r^2f^2\phi'^3-2rf\phi'^2(4f+rf') &\n\\
& +6f\phi'(f+rf'-1)-6(f-1)f'], \n & %\label{4}
%+4\ap\lambda e^{-2\phi}(f+rf'-3r^2f\phi'^2+r^2f'\phi'-1)-6\ap\lambda^2r^2e^{-4\phi}, \n
\end{flalign}
where $'$ denotes the derivative with respect to $r$. Then, the equations of motion for $f(r)$ and $\phi(r)$ read
\begin{flalign}
&r^2(rf'+f+\Lambda r^2-1)-\alpha(f-1)(2rf'-f+1)=0, &\n\\
&(r\phi'-1)^2f-1=0. &\n % -\lambda r^2e^{-2\phi}-1&=0. \n
\end{flalign}
First, the solution of $f(r)$ is
\begin{flalign}
&f(r)=1+\f{r^2}{2\alpha}\lt[1\pm\sqrt{1+4\alpha\lt(\f{2M}{r^3}+\f\Lambda 3\rt)}\rt]. &\label{5}
\end{flalign}
This is exactly the same metric solution of the 4EGB black hole. Second, the solution of $\phi(r)$ is
\begin{flalign}
&\phi(r)=\ln\f{r}{L}\pm\int_{r_+}^r\,\f{\dd u}{u\sqrt{f(u)}}, &\label{6}
%+\ln\lt[\cosh\psi(r)\pm\sqrt{1+\lambda L_0^2}\sinh\psi(r)\rt], \n
\end{flalign}
where $L$ is an integral constant. % More detailed derivation and interpretation of these solutions can be found in Ref. [91].

The black hole entropy can be related to the Noether charge associated with diffeomorphism invariance, and the Iyer--Wald formula is
\begin{flalign}
&S=-2\pi\int_\Sigma\f{\delta {\cal L}}{\delta R_{\alpha\beta\gamma\delta}}\epsilon_{\alpha\beta}\epsilon_{\gamma\delta}\,\dd\Sigma, &\n
\end{flalign}
where the integral is performed on a space-like bifurcation surface, and $\epsilon_{\alpha\beta}$ is the bivector normal to $\Sigma$. In the variation, the Riemann tensor $R_{\alpha\beta\gamma\delta}$ should be regarded as independent of metric and connection. For the static and spherically symmetric black hole metric in Eq. (\ref{5}), the Iyer--Wald formula reduces to
\begin{flalign}
&S=-8\pi\int_{r_+}\f{\delta {\cal L}}{\delta R_{trtr}}r^2\,\dd\Omega. &\label{7}
\end{flalign}
% where $r_+$ is the event horizon radius determined as the largest root of $f(r)=0$.

From Eq. (\ref{1}), we have
\begin{flalign}
&{\cal L}=\frac{1}{16\pi} \big\{R-2\Lambda+\alpha[\phi{\cal G}+4G^{\mu\nu}\partial_\mu\phi\partial_\nu\phi &\n\\
&\qquad-4(\partial\phi)^2\partial^2\phi+2((\partial\phi)^2)^2]\big\}, &\n
%&\lt. \quad -2\lambda Re^{-2\phi}-12\lambda(\p\phi)^2e^{-2\phi}-6\lambda^2e^{-4\phi}
\end{flalign}
and the variations of $R$ and ${\cal G}$ with respect to $R_{\alpha\beta\gamma\delta}$ read
\begin{flalign}
&\f{\delta R}{\delta R_{\alpha\beta\gamma\delta}}=\f{1}{2}(g^{\alpha\gamma}g^{\beta\delta}-g^{\alpha\delta}g^{\beta\gamma}), &\n\\
&\f{\delta(R_{\mu\nu}R^{\mu\nu})}{\delta R_{\alpha\beta\gamma\delta}}=\f{1}{2}(g^{\alpha\gamma}R^{\beta\delta}+g^{\beta\delta}R^{\alpha\gamma}-g^{\alpha\delta}R^{\beta\gamma} -g^{\beta\gamma}R^{\alpha\delta}), &\n\\
&\f{\delta(R_{\kappa\lambda\mu\nu}R^{\kappa\lambda\mu\nu})}{\delta R_{\alpha\beta\gamma\delta}}=2R^{\alpha\beta\gamma\delta}. &\n
\end{flalign}
Therefore, from Eq. (\ref{3}), we have
\begin{flalign}
&\f{\delta R}{\delta R_{trtr}}=-\f12, \quad \f{\delta {\cal G}}{\delta R_{trtr}}=f''. &\n
\end{flalign}
As a result, we obtain
\begin{flalign}
&\f{\delta {\cal L}}{\delta R_{trtr}}=\f{1}{16\pi}\lt(-\f 12+\alpha\phi f''\rt), &\label{8}
\end{flalign}
and other terms in ${\cal L}$, such as $G^{\mu\nu}\p_\mu\phi\p_\nu\phi$, will not contribute to the Noether charge, as already shown in Ref. \cite{94}. Substituting Eq. (\ref{8}) into Eq. (\ref{7}) and using the results in Eqs. (\ref{5}) and (\ref{6}), we finally obtain the Iyer--Wald entropy,
\begin{flalign}
&S=-8\pi\int_{r_+}\f{1}{16\pi}\lt[-\f 12+\alpha\phi(r_+) f''(r_+)\rt]r_+^2\,\dd\Omega &\n\\
&\ \ =\pi r_+^2+4\pi\alpha\ln\f{r_+}{L_0}, &\n
\end{flalign}
where $L_0$ is another integral constant different from $L$ in Eq. (\ref{6}), and the inessential integral constant from $f''(r_+)$ has been absorbed into $L_0$.

The basic reason that there is a logarithmic correction to the Bekenstein--Hawking formula lies in the fact that the scalar field $\phi$ couples to the GB term ${\cal G}$ in the way of $\alpha\phi{\cal G}$ in $S_4$. Consequently, in the integral on the event horizon, $\phi$ should be evaluated at $r_+$, and from Eq. (\ref{6}), we clearly see $\phi(r_+)=\ln (r_+/L)$. Hence, the logarithmic correction appears. Then, to determine the integral constant $L_0$, we further demand the consistency of the Smarr relation and the first law of black hole thermodynamics, and this condition requires $({\p\ln L_0}/{\p\ln\alpha})_S=1/2$. Thus, $L_0$ is fixed as $L_0=\sqrt\alpha$, and we eventually obtain the black hole entropy in the 4EGB gravity as
\begin{flalign}
&S=\pi r_+^2+4\pi\alpha\ln\f{r_+}{\sqrt\alpha}. &\n
\end{flalign}
%Altogether, the black hole entropy calculated from the action $S_4$ in Eq. (\ref{1}) via the Iyer--Wald formula reproduces the entropy calculated from the action in Eq. (\ref{2}) in the limit $d\to 4$. Actually, since these two actions lead to the same black hole solution in Eq. (\ref{5}), the same entropy is also expected.

\end{appendix}


\begin{thebibliography}{99}

\bibitem{Bekenstein}
J.D. Bekenstein, Phys. Rev. D {\bf 7}, 2333 (1973).

\bibitem{law}
J.M. Bardeen, B. Carter and S.W. Hawking, Commun. Math. Phys. {\bf 31}, 161 (1973).

\bibitem{KM}
D. Kubiz\v{n}\'{a}k and R.B. Mann, J. High Energy Phys. {\bf 07}, 033 (2012), \arxth{1205.0559}.

%\bibitem{KKK}
%D. Kastor, S. Ray, and J. Traschen, Class. Quant. Grav., {26}, 195011 (2009) %[arXiv:0904.2765 [hep-th]].

\bibitem{rev3}
D. Kubiz\v{n}\'{a}k, R.B. Mann and M. Teo, Class. Quant. Grav. {\bf 34}, 063001 (2017), \arxth{1608.06147}.

\bibitem{HP}
S.W. Hawking and D.N. Page, Commun. Math. Phys. {\bf 87}, 577 (1983).

\bibitem{Chamblin1}
A. Chamblin, R. Emparan, C.V. Johnson and R.C. Myers, Phys. Rev. D {\bf 60}, 064018 (1999), \Arxth{9902170}.

%\bibitem{Chamblin2}
%A. Chamblin, R. Emparan, C. V. Johnson et al., Phys. Rev. D, {60}, 104026 (1999) %[arXiv:hep-th/9904197 [hep-th]].

\bibitem{Witten}
E. Witten, Adv. Theor. Math. Phys. {\bf 2}, 505 (1998), \Arxth{9803131}.

%\bibitem{Herzog:2006ra}
%C. P. Herzog, Phys. Rev. Lett., {98}, 091601 (2007) %[arXiv:hep-th/0608151 [hep-th]].

\bibitem{Spallucci:2013jja}
E. Spallucci and A. Smailagic, J. Grav. {\bf 2013}, 525696 (2013), \arxth{1310.2186}.

\bibitem{Altamirano:2014tva}
N. Altamirano, D. Kubiz\v{n}\'{a}k, R.B. Mann and Z. Sherkatghanad,
%``Thermodynamics of rotating black holes and black rings: phase transitions and thermodynamic volume,''
Galaxies {\bf 2}, 89 (2014), \arxth{1401.2586}.

\bibitem{Xu:2015rfa}
J. Xu, L.-M. Cao and Y.-P. Hu,
%``P-V criticality in the extended phase space of black holes in massive gravity,''
Phys. Rev. D {\bf 91}, 124033 (2015),
%doi:10.1103/PhysRevD.91.124033
\arxgr{1506.03578}.

\bibitem{Maity:2015ida}
R. Maity, P. Roy and T. Sarkar,
%``Black Hole Phase Transitions and the Chemical Potential,''
Phys. Lett. B {\bf 765}, 386 (2017), \arxth{1512.05541}.
%doi:10.1016/j.physletb.2016.12.004[arXiv:1512.05541 [hep-th]].

\bibitem{Liu:2016uyd}
H. Liu and X.-H. Meng,
%``$P每V$ criticality in the extended phase每space of charged accelerating AdS black holes,''
Mod. Phys. Lett. A {\bf 31}, 1650199 (2016), \arxgr{1607.00496}.
%doi:10.1142/S0217732316501996 [arXiv:1607.00496 [gr-qc]].

%\bibitem{Hansen:2016ayo}
%D. Hansen, D. Kubiz\v{n}\'{a}k, and R. B. Mann,
%``Universality of P-V Criticality in Horizon Thermodynamics,''
%J. High Energy Phys., {01}, 047 (2017)
%doi:10.1007/JHEP01(2017)047 %[arXiv:1603.05689 [gr-qc]].

\bibitem{Sahay:2017hlq}
A. Sahay and R. Jha,
%``Geometry of criticality, supercriticality and Hawking-Page transitions in Gauss-Bonnet-AdS black holes,''
Phys. Rev. D {\bf 96}, 126017 (2017), \arxth{1707.03629}.
%doi:10.1103/PhysRevD.96.126017 [arXiv:1707.03629 [hep-th]].

\bibitem{Mbarek:2018bau}
S. Mbarek and R.B. Mann, J. High Energy Phys. {\bf 02}, 103 (2019), \arxth{1808.03349}.

%\bibitem{Astefanesei:2019ehu}
%D. Astefanesei, R. B. Mann, and R. Rojas,
%``Hairy Black Hole Chemistry,''
%J. High Energy Phys., {11}, 043 (2019)
%doi:10.1007/JHEP11(2019)043%[arXiv:1907.08636 [hep-th]].

\bibitem{sanren}
W. Xu, C. Wang and B. Zhu, Phys. Rev. D {\bf 99}, 044010 (2019), \arxgr{1902.10133}.

\bibitem{HPwo}
B.-Y. Su, Y.-Y. Wang and N. Li, Eur.\ Phys.\ J.\ C {\bf 80}, 305 (2020), \arxth{1905.07155}.
%``The Hawking--Page phase transitions in the extended phase space in the Gauss--Bonnet gravity,''

\bibitem{Wang:2020hjw}
P. Wang, H. Wu, H. Yang and F. Yao, J. High Energy Phys. {\bf 09}, 154 (2020), \arxgr{2006.14349}.
%arXiv:2006.14349 [gr-qc].

%%%%%%%%%%%%%%%%%%%%%%%%%%%%%%%%%%%%%%%%%%%%%%%%

\bibitem{Cai:2013qga}
R.-G. Cai, L.-M. Cao, L. Li and R.-Q. Yang,
%``P-V criticality in the extended phase space of Gauss-Bonnet black holes in AdS space,''
J. High Energy Phys. {\bf 09}, 005 (2013), \arxgr{1306.6233}.
%doi:10.1007/JHEP09(2013)005[arXiv:1306.6233 [gr-qc]].

\bibitem{Xu:2013zea}
W. Xu, H. Xu and L. Zhao,
%``Gauss-Bonnet coupling constant as a free thermodynamical variable and the associated criticality,''
Eur. Phys. J. C {\bf 74}, 2970 (2014), \arxgr{1311.3053}.
%doi:10.1140/epjc/s10052-014-2970-8 [arXiv:1311.3053 [gr-qc]].

\bibitem{Wei:2014hba}
S.-W. Wei and Y.-X. Liu,
%``Triple points and phase diagrams in the extended phase space of charged Gauss-Bonnet black holes in AdS space,''
Phys. Rev. D {\bf 90}, 044057 (2014), \arxth{1402.2837}.
%doi:10.1103/PhysRevD.90.044057 [arXiv:1402.2837 [hep-th]].

\bibitem{Mo:2014mba}
J.-X. Mo and W.-B. Liu,
%``Ehrenfest scheme for $P-V$ criticality of higher dimensional charged black holes, rotating black holes and Gauss-Bonnet AdS black holes,''
Phys. Rev. D {\bf 89}, 084057 (2014), \arxgr{1404.3872}.
%doi:10.1103/PhysRevD.89.084057 [arXiv:1404.3872 [gr-qc]].

\bibitem{Zou:2014mha}
D.-C. Zou, Y. Liu and B. Wang,
%``Critical behavior of charged Gauss-Bonnet AdS black holes in the grand canonical ensemble,''
Phys. Rev. D {\bf 90}, 044063 (2014), \arxth{1404.5194}.
%doi:10.1103/PhysRevD.90.044063 [arXiv:1404.5194 [hep-th]].

\bibitem{Belhaj:2014eha}
A. Belhaj et al., %, M. Chabab, H. El moumni et al., % K. Masmar and M. B. Sedra,
%``Maxwell's equal-area law for Gauss-Bonnet-Anti-de Sitter black holes,''
Eur. Phys. J. C {\bf 75}, 71 (2015), \arxth{1412.2162}.
%doi:10.1140/epjc/s10052-015-3299-7 [arXiv:1412.2162 [hep-th]].

\bibitem{Hendi:2016njy}
S.H. Hendi et al., % S. Panahiyan, B. Eslam Panah et al., % M. Faizal and M. Momennia,
%``Critical behavior of charged black holes in Gauss-Bonnet gravity＊s rainbow,''
Phys. Rev. D {\bf 94}, 024028 (2016), \arxgr{1607.06663}.
%doi:10.1103/PhysRevD.94.024028 [arXiv:1607.06663 [gr-qc]].

%\bibitem{Hennigar:2016ekz}
%R. A. Hennigar, E. Tjoa, and R. B. Mann,
%``Thermodynamics of hairy black holes in Lovelock gravity,''
%J. High Energy Phys., {02}, 070 (2017)
%doi:10.1007/JHEP02(2017)070[arXiv:1612.06852 [hep-th]].

\bibitem{Ghaffarnejad:2018tpr}
H. Ghaffarnejad and E. Yaraie,
%``Effects of a cloud of strings on the extended phase space of Einstein每Gauss每Bonnet AdS black holes,''
Phys. Lett. B {\bf 785}, 105 (2018), \arxgr{1806.06687}.
%doi:10.1016/j.physletb.2018.08.017 [arXiv:1806.06687 [gr-qc]].

\bibitem{Mahish:2019tgv}
S. Mahish and C. Bhamidipati,
%``Chaos in Charged Gauss-Bonnet AdS Black Holes in Extended Phase Space,''
Phys. Rev. D {\bf 99}, 106012 (2019), \arxth{1902.08932}.
%doi:10.1103/PhysRevD.99.106012 [arXiv:1902.08932 [hep-th]].

\bibitem{Zeng:2019hux}
X.-X. Zeng, X.-Y. Hu and K.-J. He,
%``Weak cosmic censorship conjecture with pressure and volume in the Gauss-Bonnet AdS black hole,''
Nucl. Phys. B {\bf 949}, 114823 (2019), \arxth{1905.07750}.
%doi:10.1016/j.nuclphysb.2019.114823 [arXiv:1905.07750 [hep-th]].

%\bibitem{Nam:2019clw}
%C. H. Nam,
%``Extended phase space thermodynamics of regular charged AdS black hole in Gauss每Bonnet gravity,''
%Gen. Rel. Grav., {51}, 100 (2019)

\bibitem{Hyun:2019gfz}
S. Hyun and C. H. Nam,
%``Charged AdS black holes in Gauss每Bonnet gravity and nonlinear electrodynamics,''
Eur. Phys. J. C {\bf 79}, 737 (2019), \arxgr{1908.09294}.
%doi:10.1140/epjc/s10052-019-7248-8 [arXiv:1908.09294 [gr-qc]].

\bibitem{Wei:2019ctz}
S.-W. Wei and Y.-X. Liu,
%``Intriguing microstructures of five-dimensional neutral Gauss-Bonnet AdS black hole,''
Phys. Lett. B {\bf 803}, 135287 (2020), \arxgr{1910.04528}.
%doi:10.1016/j.physletb.2020.135287 [arXiv:1910.04528 [gr-qc]].

\bibitem{Ghosh:2019pwy}
A. Ghosh and C. Bhamidipati,
%``Thermodynamic geometry for charged Gauss-Bonnet black holes in AdS spacetimes,''
Phys. Rev. D {\bf 101}, 046005 (2020), \arxgr{1911.06280}.
%doi:10.1103/PhysRevD.101.046005 [arXiv:1911.06280 [gr-qc]].

\bibitem{Haroon:2020vpr}
S. Haroon, R.A. Hennigar, R.B. Mann and F. Simovic,
%``Thermodynamics of Gauss-Bonnet-de Sitter Black Holes,''
Phys. Rev. D {\bf 101}, 084051 (2020), \arxgr{2002.01567}.
%doi:10.1103/PhysRevD.101.084051 [arXiv:2002.01567 [gr-qc]].

\bibitem{Ye:2020aln}
R. Ye, J. Zheng, J. Chen and Y. Wang,
%``$P每v$ criticality and heat engine efficiency for Bardeen Einstein每Gauss每Bonnet AdS black hole,''
Commun. Theor. Phys. {\bf 72}, 035401 (2020).

%%%%%%%%%%%%%%%%%%%%%%%%%%%%%%%%%%%%%%%%

%\bibitem{Lovelock}
%D. Lovelock, J. Math. Phys., {12}, 498 (1971)

\bibitem{4EGB}
D. Glavan and C. Lin, %``Einstein-Gauss-Bonnet Gravity in Four-Dimensional Spacetime,''
Phys. Rev. Lett. {\bf 124}, 081301 (2020), \arxgr{1905.03601}. %[arXiv:1905.03601 [gr-qc]].

\bibitem{Fernandes}
P.G.S. Fernandes, %``Charged Black Holes in AdS Spaces in $4D$ Einstein Gauss-Bonnet Gravity,''
Phys. Lett. B {\bf 805}, 135468 (2020), \arxgr{2003.05491}. %[arXiv:2003.05491 [gr-qc]].

\bibitem{Konoplya:2020qqh}
R.A. Konoplya and A. Zhidenko,
%``Black holes in the four-dimensional Einstein-Lovelock gravity,''
Phys. Rev. D {\bf 101}, 084038 (2020), \arxgr{2003.07788}.
%doi:10.1103/PhysRevD.101.084038 [arXiv:2003.07788 [gr-qc]].

\bibitem{Casalino:2020kbt}
A. Casalino, A. Coll\'{e}aux, M. Rinaldi and S. Vicentini, \arxgr{2003.07068}.
%``Regularized Lovelock gravity,''
%arXiv:2003.07068 [gr-qc], \arxgr{2003.07068}.

\bibitem{Kumar:2020owy}
R. Kumar and S.G. Ghosh,
%``Rotating black holes in $4D$ Einstein-Gauss-Bonnet gravity and its shadow,''
J. Cosmol. Astropart. Phys. {\bf 20}, 053 (2020), \arxgr{2003.08927}.
%doi:10.1088/1475-7516/2020/07/053 [arXiv:2003.08927 [gr-qc]].

\bibitem{Ghosh:2020syx}
S.G. Ghosh and R. Kumar,
%``Generating black holes in the novel $4D$ Einstein-Gauss-Bonnet gravity,''
J. Cosmol. Astropart. Phys. {\bf 37}, 245008 (2020), \arxgr{2003.12291}.
%arXiv:2003.12291 [gr-qc].

%\bibitem{Kumar:2020uyz}
%A. Kumar and R. Kumar,
%``Bardeen black holes in the novel $4D$ Einstein-Gauss-Bonnet gravity,''
%arXiv:2003.13104 [gr-qc]

\bibitem{Fernandes:2020nbq}
P.G.S. Fernandes, P. Carrilho, T. Clifton and D.J. Mulryne,
%``Derivation of Regularized Field Equations for the Einstein-Gauss-Bonnet Theory in Four Dimensions,''
Phys. Rev. D {\bf 102}, 024025 (2020), \arxgr{2004.08362}.
%doi:10.1103/PhysRevD.102.024025 [arXiv:2004.08362 [gr-qc]].

%\bibitem{Jusufi:2020yus}
%K. Jusufi, A. Banerjee, and S. G. Ghosh,
%``Wormholes in 4D Einstein-Gauss-Bonnet Gravity,''
%arXiv:2004.10750 [gr-qc]

\bibitem{Yang:2020jno}
K. Yang, B.-M. Gu, S.-W. Wei and Y.-X. Liu,
%``Born-Infeld Black Holes in 4D Einstein-Gauss-Bonnet Gravity,''
Eur. Phys. J. C {\bf 80}, 662 (2020), \arxgr{2004.14468}.
%doi:10.1140/epjc/s10052-020-8246-6 [arXiv:2004.14468 [gr-qc]].

\bibitem{Liu:2020yhu}
P. Liu, C. Niu, X. Wang and C.-Y. Zhang, \arxgr{2004.14267}.
%``Traversable Thin-shell Wormhole in the Novel 4D Einstein-Gauss-Bonnet Theory,''


\bibitem{Doneva:2020ped}
D.D. Doneva and S.S. Yazadjiev, \arxgr{2003.10284}.
%``Relativistic stars in 4D Einstein-Gauss-Bonnet gravity,''


\bibitem{Banerjee:2020yhu}
A. Banerjee, T. Tangphati and P. Channuie, \arxgr{2006.00479}.
%``Strange Quark Stars in 4D Einstein-Gauss-Bonnet Gravity,''


\bibitem{Islam:2020xmy}
S.U. Islam, R. Kumar and S.G. Ghosh, J. Cosmol. Astropart. Phys. {\bf 09}, 030 (2020), \arxgr{2004.01038}.
%``Gravitational lensing by black holes in $4D$ Einstein-Gauss-Bonnet gravity,''
%arXiv:2004.01038 [gr-qc].

\bibitem{Jin:2020emq}
X.-H. Jin, Y.-X. Gao and D.-J. Liu, Int. J. Mod. Phys. D {\bf 29}, 2050065 (2020), \arxgr{2004.02261}.
%``Strong gravitational lensing of a 4-dimensional Einstein-Gauss-Bonnet black hole in homogeneous plasma,''
%arXiv:2004.02261 [gr-qc].

%\bibitem{Kumar:2020sag}
%R. Kumar, S. U. Islam, and S. G. Ghosh,
%``Gravitational lensing by Charged black hole in regularized $4D$ Einstein-Gauss-Bonnet gravity,''
%arXiv:2004.12970 [gr-qc]

\bibitem{Konoplya:2020juj}
R.A. Konoplya and A. Zhidenko, Phys. Dark Univ. {\bf 30}, 100697 (2020), \arxgr{2003.12492}.
%``(In)stability of black holes in the 4D Einstein-Gauss-Bonnet and Einstein-Lovelock gravities,''
%arXiv:2003.12492 [gr-qc].

\bibitem{Zhang:2020sjh}
C.-Y. Zhang, S.-J. Zhang, P.-C. Li and M. Guo, J. High Energy Phys. {\bf 08}, 105 (2020), \arxgr{2004.03141}.
%``Superradiance and stability of the novel 4D charged Einstein-Gauss-Bonnet black hole,''
%arXiv:2004.03141 [gr-qc].

\bibitem{Liu:2020evp}
P. Liu, C. Niu and C.-Y. Zhang, \arxgr{2004.10620}.
%``Instability of the novel 4D charged Einstein-Gauss-Bonnet de-Sitter black hole,''
%arXiv:2004.10620 [gr-qc], \arxgr{2004.10620}.

\bibitem{Ge:2020tid}
X.-H. Ge and S.-J. Sin, Eur. Phys. J. C {\bf 80}, 695 (2020), \arxth{2004.12191}.
%``Causality of black holes in 4-dimensional Einstein-Gauss-Bonnet-Maxwell theory,''
%arXiv:2004.12191 [hep-th].

%\bibitem{Dadhich:2020ukj}
%N. Dadhich,
%``On causal structure of $4D$-Einstein-Gauss-Bonnet black hole,''
%arXiv:2005.05757 [gr-qc]

\bibitem{Zhang:2020qam}
C.-Y. Zhang, P.-C. Li and M. Guo, Eur. Phys. J. C {\bf 80}, 874 (2020), \arxth{2003.13068}.
%``Greybody factor and power spectra of the Hawking radiation in the novel $4D$ Einstein-Gauss-Bonnet de-Sitter gravity,''
%arXiv:2003.13068 [hep-th].

\bibitem{Liu:2020vkh}
C. Liu, T. Zhu and Q. Wu,
%``Thin Accretion Disk around a four-dimensional Einstein-Gauss-Bonnet Black Hole,''
Chin. Phys. C {\bf 45}, 015105 (2020), \arxgr{2004.01662}.
%arXiv:2004.01662 [gr-qc].

\bibitem{Yang:2020czk}
S.-J. Yang et al., % J.-J. Wan, J. Chen et al., % J. Yang and Y. Q. Wang,
%``Weak cosmic censorship conjecture for the novel $4D$ charged Einstein-Gauss-Bonnet black hole with test scalar field and particle,''
Eur. Phys. J. C {\bf 80}, 937 (2020), \arxgr{2004.07934}.
%arXiv:2004.07934 [gr-qc].

\bibitem{Ying:2020bch}
S. Ying,
%``Thermodynamics and Weak Cosmic Censorship Conjecture of 4D Gauss-Bonnet-Maxwell Black Holes via Charged Particle Absorption,''
Chin. Phys. C {\bf 44}, 125101 (2020), \arxgr{2004.09480}.
%arXiv:2004.09480 [gr-qc].

\bibitem{Konoplya:2020bxa}
R.A. Konoplya and A.F. Zinhailo,
%``Quasinormal modes, stability and shadows of a black hole in the novel 4D Einstein-Gauss-Bonnet gravity,''
Eur. Phys. J. C {\bf 80}, 1049 (2020), \arxgr{2003.01188}.
%arXiv:2003.01188 [gr-qc].

\bibitem{Aragon:2020qdc}
A. Arag\'{o}n, R. B\'{e}car, P.A. Gonz\'{a}lez and Y. V\'{a}squez, Eur. Phys. J. C {\bf 80}, 773 (2020), \arxgr{2004.05632}.
%``Perturbative and nonperturbative quasinormal modes of 4D Einstein-Gauss-Bonnet black holes,''
%arXiv:2004.05632 [gr-qc].

\bibitem{Guo:2020zmf}
M. Guo and P.-C. Li,
%``Innermost stable circular orbit and shadow of the $4D$ Einstein每Gauss每Bonnet black hole,''
Eur. Phys. J. C {\bf 80}, 588 (2020), \arxgr{2003.02523}.
%doi:10.1140/epjc/s10052-020-8164-7 [arXiv:2003.02523 [gr-qc]].

\bibitem{Wei:2020ght}
S.-W. Wei and Y.-X. Liu, \arxgr{2003.07769}.
%``Testing the nature of Gauss-Bonnet gravity by four-dimensional rotating black hole shadow,''
%arXiv:2003.07769 [gr-qc], \arxgr{2003.07769}.

\bibitem{Zeng:2020dco}
X.-X. Zeng, H.-Q. Zhang and H. Zhang, Eur. Phys. J. C {\bf 80}, 872 (2020), \arxgr{2004.12074}.
%``Shadows and photon spheres with spherical accretions in the four-dimensional Gauss-Bonnet black hole,''
%arXiv:2004.12074 [gr-qc].

\bibitem{Hegde:2020xlv}
K. Hegde et al., \arxgr{2003.07769}. % A. Naveena Kumara, C. L. A. Rizwan et al., % A. K. M. and M. S. Ali,
%``Thermodynamics, Phase Transition and Joule Thomson Expansion of novel 4-D Gauss Bonnet AdS Black Hole,''
%arXiv:2003.08778 [gr-qc], \arxgr{2003.07769}.

\bibitem{Singh:2020xju}
D. V. Singh and S. Siwach, Phys. Lett. B {\bf 808}, 135658 (2020), \arxgr{2003.11754}.
%``Thermodynamics and P-v criticality of Bardeen-AdS Black Hole in 4-D Einstein-Gauss-Bonnet Gravity,''
%^arXiv:2003.11754 [gr-qc].

\bibitem{HosseiniMansoori:2020yfj}
S. A. Hosseini Mansoori, \arxgr{2003.13382}.
%arXiv:2003.13382 [gr-qc], \arxgr{2003.13382}.
%``Thermodynamic geometry of the novel 4-D Gauss Bonnet AdS Black Hole,''

\bibitem{Wei:2020poh}
S.-W. Wei and Y.-X. Liu,
%``Extended thermodynamics and microstructures of four-dimensional charged Gauss-Bonnet black hole in AdS space,''
Phys. Rev. D {\bf 101}, 104018 (2020), \arxgr{2003.14275}.
%doi:10.1103/PhysRevD.101.104018 [arXiv:2003.14275 [gr-qc]].

\bibitem{EslamPanah:2020hoj}
B. Eslam Panah, K. Jafarzade and S.H. Hendi,
Nucl. Phys. B {\bf 961}, 115269 (2020), \arxth{2004.04058}.
%arXiv:2004.04058 [hep-th].
%``Charged 4D Einstein-Gauss-Bonnet-AdS Black Holes: Shadow, Energy Emission, Deflection Angle and Heat Engine,''

\bibitem{Qiao:2020hkx}
X. Qiao et al., \arxth{2005.01007}. % L. OuYang, D. Wang et al., % Q. Pan and J. Jing,
%``Holographic superconductors in 4D Einstein-Gauss-Bonnet gravity,''
%arXiv:2005.01007 [hep-th], \arxth{2005.01007}.

%\bibitem{Singh:2020mty}
%D. V. Singh, R. Kumar, S. G. Ghosh et al., % and S. D. Maharaj,
%``Phase transition of AdS black holes in 4D EGB gravity coupled to nonlinear electrodynamics,''
%arXiv:2006.00594 [gr-qc]

\bibitem{Li:2020vpo}
H.-L. Li, X.-X. Zeng and R. Lin,
%``Holographic phase transition from novel Gauss每Bonnet AdS black holes,''
Eur. Phys. J. C {\bf 80}, 652 (2020).

\bibitem{Odintsov:2020zkl}
S. Odintsov and V. Oikonomou,
%``Swampland Implications of GW170817-compatible Einstein-Gauss-BonnetGravity,''
Phys. Lett. B {\bf 805}, 135437 (2020), \arxgr{2004.00479}. %doi:10.1016/j.physletb.2020.135437[arXiv:2004.00479 [gr-qc]].

\bibitem{Odintsov:2020sqy}
S. Odintsov, V. Oikonomou and F. Fronimos, Nucl. Phys. B {\bf 958}, 115135 (2020), \arxgr{2003.13724}. %arXiv:2003.13724 [gr-qc].
%``Rectifying Einstein-Gauss-Bonnet Inflation in View of GW170817,''

\bibitem{Odintsov:2020xji}
S.D. Odintsov, V.K. Oikonomou and F.P. Fronimos, Annals Phys. {\bf 420}, 168250 (2020), \arxgr{2007.02309}. %arXiv:2007.02309 [gr-qc].
%``Non-Minimally Coupled Einstein Gauss Bonnet Inflation Phenomenology in View of GW170817,''

\bibitem{Oikonomou:2020oil}
V.K. Oikonomou and F.P. Fronimos, Europhys. Lett. {\bf 131}, 30001 (2020), \arxgr{2007.11915}. %arXiv:2007.11915 [gr-qc].
%``A Nearly Massless Graviton in Einstein-Gauss-Bonnet Inflation withLinear Coupling Implies Constant-roll for the Scalar Field,''

\bibitem{Li:2020tlo}
S.-L. Li, P. Wu and H. Yu, \arxgr{2004.02080}.
%``Stability of the Einstein Static Universe in $4 D$ Gauss-Bonnet Gravity,''
%arXiv:2004.02080 [gr-qc], \arxgr{2004.02080}.

\bibitem{Haghani:2020ynl}
Z. Haghani, Phys. Dark Univ. {\bf 30}, 100720 (2020), \arxgr{2005.01636}.
%``Growth of matter density perturbations in 4D Einstein-Gauss-Bonnet gravity,''
%arXiv:2005.01636 [gr-qc].

\bibitem{Narain:2020qhh}
G. Narain and H.-Q. Zhang, \arxgr{2005.05183}.
%``Cosmic evolution in novel-Gauss Bonnet Gravity,''
%arXiv:2005.05183 [gr-qc], \arxgr{2005.05183}.

\bibitem{Aoki:2020iwm}
K. Aoki, M.A. Gorji and S. Mukohyama, J. Cosmol. Astropart. Phys. {\bf 09}, 014 (2020), \arxgr{2005.08428}.
%``Cosmology and gravitational waves in consistent $D\to 4$ Einstein-Gauss-Bonnet gravity,''
%arXiv:2005.08428 [gr-qc].

\bibitem{MohseniSadjadi:2020jmc}
H. Mohseni Sadjadi, \arxgr{2005.10024}.
%``On cosmic acceleration in four dimensional Einstein-Gauss-Bonnet gravity,''

\bibitem{Clifton:2020xhc}
T. Clifton, P. Carrilho, P.G.S. Fernandes and D.J. Mulryne, Phys. Rev. D {\bf 102}, 084005 (2020), \arxgr{2006.15017}.
%``Observational Constraints on the Regularized 4D Einstein-Gauss-Bonnet Theory of Gravity,''
%arXiv:2006.15017 [gr-qc].

\bibitem{Feng:2020duo}
J.-X. Feng, B.-M. Gu and F.-W. Shu, \arxgr{2006.16751}.
%``Theoretical and observational constraints on regularized 4$D$ Einstein-Gauss-Bonnet gravity,''
%arXiv:2006.16751 [gr-qc], \arxgr{2006.16751}.

\bibitem{Garcia-Aspeitia:2020uwq}
M. A. Garc\'{\i}a-Aspeitia and A. Hern\'{a}ndez-Almada, \arxas{2007.06730}.
%``Einstein-Gauss-Bonnet gravity: constraining with current cosmological observations,''
%arXiv:2007.06730 [astro-ph.CO], \arxas{2007.06730}.

%%%%%%%%%%%%%%%%%%%%%%%%%%%%%%%

\bibitem{Ai:2020peo}
W.-Y. Ai, Commun. Theor. Phys. {\bf 72}, 095402 (2020), \arxgr{2004.02858}.
%``A note on the novel 4D Einstein-Gauss-Bonnet gravity,'' %doi:10.1088/1572-9494/aba242 [arXiv:2004.02858 [gr-qc]].

\bibitem{Gurses:2020ofy}
M. G\"{u}rses, T.\c{C}. \c{S}i\c{s}man and B. Tekin,
%``Is there a novel Einstein-Gauss-Bonnet theory in four dimensions?,''
Eur. Phys. J. C {\bf 80}, 647 (2020), \arxgr{2004.03390}.
%doi:10.1140/epjc/s10052-020-8200-7 [arXiv:2004.03390 [gr-qc]].

\bibitem{Mahapatra:2020rds}
S. Mahapatra,
Eur. Phys. J. C {\bf 80}, 992 (2020), \arxgr{2004.09214}.
%``A note on the total action of $4D$ Gauss-Bonnet theory,''
%arXiv:2004.09214 [gr-qc].

\bibitem{Shu:2020cjw}
F.-W. Shu,
Phys. Lett. B {\bf 811}, 135907 (2020), \arxgr{2004.09339}.
%``Vacua in novel 4D Einstein-Gauss-Bonnet Gravity: pathology and instability?,''
%arXiv:2004.09339 [gr-qc].

\bibitem{Bonifacio:2020vbk}
J. Bonifacio, K. Hinterbichler and L.A. Johnson,
%``Amplitudes and 4D Gauss-Bonnet Theory,''
Phys. Rev. D {\bf 102}, 024029 (2020), \arxth{2004.10716}. %doi:10.1103/PhysRevD.102.024029 [arXiv:2004.10716 [hep-th]].

\bibitem{Hennigar:2020lsl}
R.A. Hennigar, D. Kubiz\v{n}\'{a}k, R.B. Mann and C. Pollack,
%``On taking the D ↙ 4 limit of Gauss-Bonnet gravity: theory and solutions,''
J. High Energy Phys. {\bf 07}, 027 (2020), \arxgr{2004.09472}.
%doi:10.1007/JHEP07(2020)027 [arXiv:2004.09472 [gr-qc]].

\bibitem{Tian:2020nzb}
S. X. Tian and Z.-H. Zhu, \arxgr{2004.09954}.
%``Comment on "Einstein-Gauss-Bonnet Gravity in Four-Dimensional Spacetime",''
%arXiv:2004.09954 [gr-qc], \arxgr{2004.09954}.

\bibitem{Arrechea:2020evj}
J. Arrechea, A. Delhom and A. Jim\'{e}nez-Cano, \arxgr{2004.12998}.
%``Yet another comment on four-dimensional Einstein-Gauss-Bonnet gravity,''
%arXiv:2004.12998 [gr-qc], \arxgr{2004.12998}.

%%%%%%%%%%%%%%%%%%%%%%%%%%%%

\bibitem{Aoki:2020lig}
K. Aoki, M.A. Gorji and S. Mukohyama,
Phys. Lett. B {\bf 810}, 135843 (2020), \arxgr{2005.03859}.
%``A consistent theory of $D\rightarrow 4$ Einstein-Gauss-Bonnet gravity,''
%arXiv:2005.03859 [gr-qc].

\bibitem{Easson:2020mpq}
D.A. Easson, T. Manton and A. Svesko,
J. Cosmol. Astropart. Phys. {\bf 10}, 026 (2020), \arxth{2005.12292}.
%``$D\to4$ Einstein-Gauss-Bonnet Gravity and Beyond,''
%arXiv:2005.12292 [hep-th].

\bibitem{Lin:2020kqe}
Z.-C. Lin et al., % K. Yang, S.-W. Wei et al., % Y. Q. Wang and Y. X. Liu,
Eur. Phys. J. C {\bf 80}, 1033 (2020), \arxgr{2006.07913}.
%``Is the four-dimensional novel EGB theory equivalent to its regularized counterpart in a cylindrically symmetric spacetime?,''
%arXiv:2006.07913 [gr-qc].

%%%%%%%%%%%%%%%%%%%%%%%%%%%%%%%%%%%%

\bibitem{Boulware:1985wk}
D.G. Boulware and S. Deser, Phys. Rev. Lett. {\bf 55}, 2656 (1985).

%\bibitem{Smarr}
%L. Smarr,
%``Mass formula for Kerr black holes,''
%Phys. Rev. Lett., {30}, 71 (1973)

\bibitem{caii}
R.-G. Cai, L.-M. Cao and N. Ohta, %\Black Holes in Gravity with Conformal Anomaly and Logarithmic Term in Black Hole Entropy,"
J. High Eenergy Phys. {\bf 04}, 082 (2010), \arxth{0911.4379}.
%doi:10.1007/JHEP04(2010)082 [arXiv:0911.4379 [hep-th]].

\bibitem{lue}
H. L\"{u} and Y. Pang, Phys. Lett. B {\bf 809}, 135717 (2020), \arxgr{2003.11552}. %``Horndeski Gravity as $D\rightarrow4$ Limit of Gauss-Bonnet,''
%arXiv:2003.11552 [gr-qc].

\bibitem{Wald}
V. Iyer and R.M. Wald,
%``Some properties of Noether charge and a proposal for dynamical black hole entropy,''
Phys. Rev. D {\bf 50}, 846 (1994), \Arxgr{9403028}. %[arXiv:gr-qc/9403028 [gr-qc]].

\bibitem{Kubiznak:2014zwa}
D. Kubiz\v{n}\'{a}k and R.B. Mann,
%``Black hole chemistry,''
Can. J. Phys. {\bf 93}, 999 (2015), \arxgr{1404.2126}. % [arXiv:1404.2126 [gr-qc]].

\bibitem{94}
X.-H. Feng, H.-S. Liu, H. L\"{u} and C.N. Pope, J. High Energy Phys. {\bf 11}, 176 (2015), \arxth{1509.07142}.
\end{thebibliography}
\end{document}